\documentclass[journal,12pt,draftclsnofoot,onecolumn]{IEEEtran}
\IEEEoverridecommandlockouts
\usepackage[utf8]{inputenc}
\usepackage[T1]{fontenc}
\usepackage{url}
\usepackage{ifthen}
\usepackage{cite}
\usepackage{amsmath} 
\usepackage{textcomp}
\usepackage{soul,color,xcolor}
\usepackage{verbatim}
\usepackage{bm}
\usepackage{setspace}
\usepackage{algorithmic}
\usepackage[ruled]{algorithm2e}                      
\usepackage{amsthm}
\usepackage{amsfonts}
\usepackage{mathrsfs}
\usepackage{multirow}
\usepackage{citesort}
\usepackage{subfigure}
\usepackage{bm}
\usepackage{adjustbox}
\usepackage{graphicx}
\usepackage{epstopdf}
\usepackage{amssymb}
\usepackage{diagbox}
\usepackage{makecell}
\usepackage{caption}

\usepackage{xhfill}
\allowdisplaybreaks[4]
\newcommand{\xfilll}[2][1ex]{%
	\dimen0=#2\advance\dimen0 by #1%
	\leaders\hrule height \dimen0 depth -#1\hfill%
}

\newenvironment{Proof}{
	{\indent\it Proof:}\;}{\hfill $\blacksquare$\par}
\newtheoremstyle{thry}
{3pt}
{3pt}
{}
{1em}
{}
{:}
{.5em}
{}
\theoremstyle{thry}

\newtheorem{theorem}{{\textbf{Theorem}}}
\newtheorem{definition}{{\textbf{Definition}}}
\newtheorem{lemma}{{\textbf{Lemma}}}
\newtheorem{corollary}{{\textbf{Corollary}}}
\newtheorem{example}{\textbf{Example}}
\definecolor{mycolor}{rgb}{0.0, 0.5, 1.0}

\begin{document}
\title{Systematic Bernoulli Generator Matrix Codes\\
\thanks{ This work was supported by the National Key R\&D Program of China~(Grant No.~2021YFA1000500). Part of this work was presented at 2022 International Symposium on Information Theory~[40]. (\textit{Corresponding author: Xiao Ma.} )}
}

\author{
  \IEEEauthorblockN{   Yixin Wang, Fanhui Meng and  Xiao Ma}\\
  \thanks{ Yixin Wang and Fanhui Meng are with the School of Systems Science and Engineering, Sun Yat-sen University,  Guangzhou 510006, China~(e-mail:
 	wangyx58@mail2.sysu.edu.cn, mengfh3@mail2.sysu.edu.cn)}
 \thanks{  Xiao Ma is  with the Guangdong Key Laboratory of Information Security Technology, School of Computer Science and Engineering, Sun Yat-sen University,  Guangzhou 510006, China~(e-mail:
 	maxiao@mail.sysu.edu.cn)}

}

\maketitle

\begin{abstract}
This paper is concerned with the systematic Bernoulli generator matrix~(BGM) codes, which have been proved to be capacity-achieving over
binary-input output-symmetric~(BIOS) channels in terms of bit-error rate~(BER). We prove that the systematic BGM codes are also capacity-achieving over BIOS channels in terms of frame-error rate (FER). To this end, we present a new framework to prove the coding theorems for binary linear codes. Different from the widely-accepted approach via ensemble enlargement, the proof directly applies to the systematic binary linear codes. The new proof indicates that the pair-wise independence condition is not necessary for proving the binary linear code ensemble to achieve the capacity of the BIOS channel. The Bernoulli parity-check~(BPC) codes, which fall within the framework of the systematic BGM codes with parity-check bits known at the decoder can also be proved to achieve the capacity. The presented framework also reveals a new mechanism pertained to the systematic linear codes that  the systematic bits and the corresponding parity-check bits play different roles. Precisely, the noisy systematic bits are used to limit the list size of candidate codewords, while the noisy parity-check bits are used to select from the list the maximum likelihood codeword.  For systematic BGM codes with finite length, we derive the lower bounds on the BER and FER, which can be used to predict the error floors. Numerical results show that the systematic BGM codes match well with the derived error floors. The performance in water-fall region can be improved with approaches in statistical physics and the error floors can be significantly improved by implementing the concatenated codes with the systematic BGM codes as the inner codes.  
\end{abstract}

\begin{IEEEkeywords}
Coding theorem,  linear codes, low density generator matrix~(LDGM) codes, low density parity-check~(LDPC) codes, partial error exponent, partial mutual information, systematic Bernoulli generator matrix~(BGM) codes
\end{IEEEkeywords}

\section{Introduction}

\par Linear codes play an important role in the channel coding theory~\cite{Gallager1968}.  It was first proved in~\cite{Elias1955Coding} that the totally random linear code ensemble can achieve the capacity of binary symmetric channels~(BSCs). The same conclusion was drawn in~\cite[Theorem 6.2.1]{Gallager1968} by deriving the error exponent. Many modern capacity-approaching/achieving linear codes have been proposed, including the low density parity-check (LDPC) codes, which were invented by Gallager in the early 1960s~\cite{1962Gallager}.  It has been proved~(numerically by density evolution~(DE)) in~\cite{Richardson2001DE,Richardson2001LDPC} 
 that LDPC codes can be a class of capacity-approaching under iterative message passing decoding over a broad class of channels. In contrast, the dual of LDPC codes, namely, the low density generator matrix~(LDGM) codes, are not that good at large as recognized by Mackay during his rediscovery of LDPC codes~\cite{MacKay1999sparse}. As an exceptional work, Sourlas built a bridge from error correcting codes to spin glass in~1989~\cite{sourlas1989spin} and demonstrated that an extremely low-rate regular LDGM code can be good for binary phase shift keying~(BPSK) signaling over additive white Gaussian noise~(AWGN) channels. More general regular LDGM codes were investigated in~\cite{cheng1996some}. Similar to LDPC codes, conventional LDGM code ensembles are usually characterized by degree distribution polynomials with constraints on maximum degrees.  Such constraints inevitably introduce dependence between the columns and the rows of the generator matrices and bring inconvenience for performance analysis. Partially for this reason, several classes of LDGM codes have been analyzed by generating randomly and independently each column of the generator matrices according to certain distributions  without maximum degree constraints. For example,  Luby transform~(LT) codes~\cite{Luby2002LT} can be viewed as rateless  LDGM codes, where each coded symbol is generated independently controlled by the ideal Soliton distribution or the robust Soliton distribution.  Another LDGM code ensemble without maximum degree constraint was analyzed in~\cite{Montanari2005Tight}, where the degree of the variable nodes follows a Poisson distribution. The performance of LDGM codes are not that good as channel codes due to their higher error floors caused by the low weight codewords. However, the error floors can be lowered down by concatenating with high-rate outer codes~\cite{2003Approaching,Lopez2007Serially}. It has been shown that Raptor codes~(concatenation of outer linear block codes and inner LT codes)~\cite{Shokrollahi2006Raptor} can achieve the capacity of the binary erasure channels~(BECs). 
{As a class of sparse codes with efficient message-passing algorithms, the LDGM codes have a wide range of applications~\cite{Wainwright2007}, including  source codes\cite{Zhu2021Compression}, lossy  compression codes~\cite{Wainwright2010,Golmohammadi2018,Alinia2023}, erasure correct correcting codes~\cite{Luby2002LT,Shokrollahi2006Raptor}~(which can be improved for use in noisy channels~\cite{Etesami2006,Shirvanimoghaddam2016,Kharel2018}), error reduction codes in concatenation~\cite{Lopez2007Serially,Zhang2017} and steganographic codes~\cite{Alinia2024,Yao2024}.}


\par 

\par Systematic linear codes have the information bits in the codewords, which can benefit the encoding and decoding procedure compared with non-systematic linear codes. Of the same codeword length, systematic linear codes can have less operating steps in the coding procedure. More importantly, using systematic instead of non-systematic linear codes allows the decoder to obtain the decoded bits directly from the received sequences. However, most existing coding theorems are proved for non-systematic codes and direct proofs are rarely found for systematic codes. The systematic Bernoulli generator matrix (BGM) codes are a class of systematic linear block codes~\cite{Ma2016Coding,Cai2020SCLDGM}. It has been proved that, in terms of bit-error rate~(BER), the systematic BGM code ensembles are capacity-achieving for binary-input output-symmetric~(BIOS) memoryless channels~\cite{Ma2016Coding,Cai2020SCLDGM}. Notice that the BGM codes investigated in~\cite{Kakhaki2012LDGM} are non-systematic, where their capacity-achievability was  only proved for BSCs. In this paper,  we propose a new framework to prove the channel coding theorem for linear codes. With this framework, we  prove that the systematic BGM codes are also capacity-achieving over BIOS channels in terms of frame-error rate~(FER). The Bernoulli parity-check~(BPC) codes, which fall within the presented framework of the systematic BGM codes with parity-check bits known at the decoder can also be proved to achieve the capacity. In the presented framework, the systematic bits and the corresponding parity-check bits play different roles. Precisely, the noisy systematic bits are used to limit the list size of candidate codewords, while the noisy parity-check bits are used to select from the list the maximum likelihood codeword. For systematic BGM codes with finite length, we derive the lower bounds on the BER and FER, which can be used to predict the error floors. Numerical results show that the  systematic BGM codes  match well with the derived error floors.  To improve the water-fall region performance of systematic BGM codes, we turn to the approaches in statistical physics, while to improve the error-floor region performance, we concatenate the systematic BGM codes with outer codes. The contributions of this paper are summarized as follows.
\begin{itemize}
	\item [1)] \textbf{Coding Theorem for Systematic BGM Code Ensemble:}  We propose a new framework to prove the channel coding theorem for linear codes and  prove that the systematic BGM codes~(as a class of LDGM codes) are  capacity-achieving over BIOS channels in terms of FER with this framework.
	\item [2)] \textbf{Coding Theorem for BPC Code Ensemble:} We prove that within the presented framework, the BPC codes~(as a class of LDPC codes) are  capacity-achieving over BIOS channels.
	\item [3)] \textbf{Performance Analysis for Finite Length BGM Codes:} We derive the lower bounds on the BER and FER bounds for finite length systematic BGM codes.
	\item [4)]\textbf{Improving the Performance of BGM Codes:} We relate the BGM codes to the complex network and discover that the assortativity coefficients of the systematic BGM codes can influence the performance of systematic BGM codes. So we optimize the assortativity coefficient of the systematic BGM codes to improve the performance.
\end{itemize}

\par The rest of the paper is organized as follows. In Section~\ref{sec2}, we describe the new framework and the conventional coding theorem for binary linear codes.  In Section~\ref{sec3}, we give the main results of this paper. In Section~\ref{sec4},  we prove the coding theorems for the systematic BGM codes and BPC codes with the presented framework.   In Section~\ref{sec5}, we derive the lower bounds on the BER and FER for finite length systematic BGM codes and present simulation results. We then improve the performance of BGM codes in water-fall region and error-floor region in Section~\ref{sec6}. Section~\ref{sec7} concludes this paper.
\par In this paper, a random variable is denoted by an upper-case letter, say $X$, whose realization is denoted by the corresponding lowercase letter $x\in\mathcal{X}$. We use $P_{X}(x)$, $x\in \mathcal{X}$ to represent the probability mass~(or density) function of a random discrete~(or continuous) variable. For a vector of length $m$, we represent it as $\bm x=(x_0,x_1,\cdots,x_{m-1})$. We also use $\bm x^m$ to emphasize the length of $\bm x$. We denote by $\mathbb{F}_2=\{0,1\}$ the binary field. We denote  by $\log$  the base-2 logarithm and by $\exp$  the base-2 exponent.
\section{A New Framework}
\label{sec2}

\subsection{Problem Statement}
\par
\begin{figure}[t]
	\centering
	\includegraphics[width=0.6\textwidth]{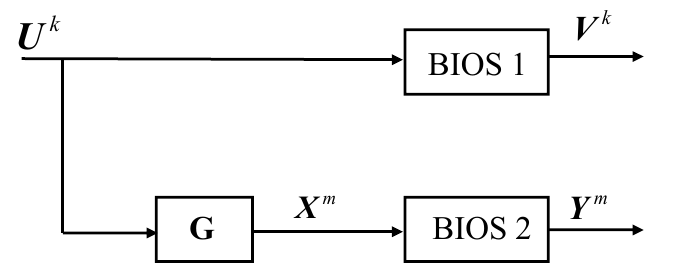}
	\caption{A  system model with systematic linear coding. }
	\label{systemm model}
\end{figure}

 We consider a system model that is depicted in Fig.~\ref{systemm model}, where $\bm U^{k}\in \mathbb{F}_2^{k}$ is a segment of a Bernoulli process  with a success probability of $\theta = P_U(1)$ and referred to as the message bits to be transmitted, ${\mathbf{G}}$ is a  binary matrix of size $k\times m$ and $\bm X^m=\bm U^k{\mathbf{G}}\in \mathbb{F}_{2}^{m}$ is referred to as parity-check bits corresponding to $\bm U^k$. The message bits $\bm U^k$ and the parity-check bits $\bm X^m$ are transmitted through two~(possibly different) BIOS channels, resulting in $\bm V^k$ and $\bm Y^m$, respectively. A  BIOS  memoryless channel is characterized by an input  $x \in \mathcal{X}=\mathbb{F}_{2}$, an output set $\mathcal{Y}$~(discrete or continuous), and a conditional probability mass (or density) function\footnote{If the context is clear, we may omit the subscript of the probability mass (or density) function.}$\{P_{Y|X}(y|x)\big| x\in\mathbb{F}_2, y\in\mathcal{Y}\}$ which satisfies the symmetric condition that $P_{Y|X}(y|1)=P_{Y|X}(\pi(y)|0)$ for some mapping $\pi: \mathcal{Y}\rightarrow\mathcal{Y}$ with $\pi^{-1}(\pi(y))=y$. For simplicity, we assume that the BIOS channels are  memoryless, meaning that $P_{\bm{V}|\bm{U}}(\bm v|\bm u)=\prod\limits_{i=0}^{k-1}P_{V|U}(v_i|u_i)$ and $P_{\bm{Y}|\bm{X}}(\bm y|\bm x)=\prod\limits_{i=0}^{m-1}P_{Y|X}(y_i|x_i)$. For a Bernoulli input $X$, we can define the mutual information $I(X; Y)$. The channel capacity of the BIOS channel is given by $C = I(X; Y)$ with $X$ being a uniform binary random variable  with $P_X(0) = P_X(1) = 1/2$.
\par  The task of the receiver is to recover $\bm U^k$  from $(\bm V^k,\bm Y^m)$.  The code rate $R$~of the considered systematic linear code is defined as $R=k/(k+m)$. The BER   is defined as $\textbf{E}[W_H(\hat{\bm U}^k+\bm U^k)]/k$, where $\mathbf{E}[\cdot]$ denotes the expectation of the random variable, $W_H(\cdot)$ denotes the Hamming weight function, and $\hat{\bm U}^k$ is the estimate of $\bm U^k$ from decoding.  The FER is defined as ${\rm Pr}\{\hat{\bm U}^k\neq \bm U^k\}$. 
\par We are primarily concerned with the following question: for sufficiently large $k$, how many parity-check bits are required for reliably transmitting $\bm U^k$? Intuitively, the reliable transmission of $\bm U^k$ can be achieved if $kH(U|V)<mI(X;Y)$. The limit of the ratio $m/k$ as $k$ goes to infinity is considered in this paper, which represents the average number of parity-check bits required per message bit for reliable transmission. The framework unifies the following problems in coding theory. 
\begin{itemize}
   \item If  BIOS~$1$ and BIOS~$2$ are the same, and $U$ is uniform, the problem is equivalent to the channel coding problem. From $kH(U|V)<mI(X;Y)$, we have { $m/k>H(U|V)/I(X;Y)$}, which is equivalent to $k/(k+m)<C$ and $k/(k+m)$ is the code rate for channel coding. This can be verified by noting that $C=1-H(X|Y)$ and $H(U|V)=H(X|Y)$.
   \item If  BIOS~$1$ is a totally erased channel\footnote{By the totally erased channel we mean an erasure channel with erased probability~$1$.} and BIOS~$2$ is noiseless, the problem is equivalent to the source coding problem. From $kH(U|V)<mI(X; Y)$, we have  $m/k>H(U)$ and $m/k$ is the code rate for the source coding. This can be verified by noting that $H(U|V)=H(U)$ and $I(X; Y)=1$.
   \item If  BIOS 1 is a totally erased channel, and BIOS~$2$ has noise, the problem can be viewed as the joint source-channel coding~(JSCC) problem for Bernoulli sources.  From $kH(U|V)<mI(X;Y)$, we have $m/k>H(U)/I(X;Y)$, where $m/k$ is the transmission ratio and the limit of the minimum transmission ratio~(LMTR) is $H(U)/C$~\cite{Csizar2011}.
   \item If BIOS~$1$ is a (possibly) noisy channel and BIOS~$2$ is a noiseless channel, the problem is equivalent to transmit the uniform sources with the parity-check codes specified by the random sparse parity-check matrix~$\mathbf{H}=\mathbf{G}^T$. From~$kH(U|V)<mI(X;Y)$, we have $m/k>H(U|V)$ for $I (X;Y) = 1$. Equivalently, we have $(k-m)/k< C$, where $(k-m)/k$ is the design code rate of the parity-check code.

   \item If BIOS~$2$ is noiseless, and the BIOS~1 is disconnected but $\bm V^k$ is received from another source as side information of $\bm U^k$, the problem can be viewed as the distributed source coding problem. From $kH(U|V)<mI(X;Y)$, we have {$m/k>H(U|V)$} for $I(X;Y)=1$, which corresponds to one corner point of the Slepian-Wolf encoding region~\cite{slepian1973correlated} .  
\end{itemize}
\par In this paper, we mainly focus on channel coding with systematic generator matrix codes or (possibly non-systematic) parity-check codes. We mainly consider in this paper the case of $\theta=P_U(1)=1/2$.

\subsection{Coding Theorem for Systematic Linear Codes}
\par We present the following coding theorem for systematic linear codes, which is well-known~(say, \cite[Theorem 6.2.1]{Gallager1968} for BSCs) but, surprisingly, has no direct proof available in the literature (to the best of our knowledge).
\begin{theorem}
	\label{general_theorem} \rm
	Consider systematic binary linear block codes defined by the generator matrices of the form $[\mathbf{I}~\mathbf{G}]$, where $\mathbf{I}$ is the identity matrix of order $k$ and $\mathbf{G}$ is a binary matrix of size $k\times m$. For two arbitrarily small positive numbers $\epsilon$ and $\delta$, one can always find a sufficiently large integer $k_0$ such that, for all $k \geq k_0$ and $m=\lfloor k(1-C+\delta)/(C-\delta)\rfloor$, there exists a matrix $\mathbf{G}$  of size $k\times m$ satisfying that  $m/k\leq (1-C+\delta)/(C-\delta)$  and the maximum likelihood decoding~(MLD) error rate ${\rm FER} \leq \epsilon$.
\end{theorem} 
\par \emph{Outline Proof of Theorem 1: }The~(indirect) proof follows from that of~\cite[Theorem 6.2.1]{Gallager1968}. The outline of the proof is to prove first the existence of capacity-achieving~(non-systematic) codes and then the generator matrices of such codes to be full-rank~(also see~\cite[Lemma~4.15]{roth2006introduction}). Then, systematic generator matrices of the form $[\mathbf{I}~\mathbf{G}]$ can be obtained by performing elementary row transformations and possibly column permutations.
\par For {$m/k\leq (1-C+\delta)/(C-\delta)$}, we have the code rate $R=k/(k+m)\geq C-\delta$. First, instead of the systematic codes, we consider an enlarged code ensemble. For this purpose, define a coset code ensemble by generating a totally random binary~(non-systematic) matrix $\mathbf{A}$ of size~$k\times (k+m)$  and a totally random vector $\bm v\in \mathbb{F}_2^{k+m}$. By ``totally random'' we mean that each component is generated independently and uniformly at random. A message vector $\bm u^k \in \mathbb{F}_2^k$ is  encoded into a codeword $\bm u \mathbf{A}+\bm v$ and then transmitted over a  BIOS  memoryless channel. With the random vector $\bm v$, the codewords are  uniformly distributed (with probability~$2^{-(k+m)}$)  over~$\mathbb{F}_2^{k+m}$. It can be further proved that the codewords are pair-wise independent due to the total randomness of $\mathbf{A}$. Consequently, we can apply~\cite[Theorem 5.6.2]{Gallager1968} to assert   the existence of a coset code with an average frame  error probability at most~$\exp [- (k+m)E_r(R)]$ under the MLD, where $E_r(R)$ is the error exponent and $R=k/(k+m)$ is the code rate of the coset code. For  BIOS   memoryless channels, the linear  code defined by the generator matrix  $\mathbf{A}$~(by removing the coset representative $\bm v$) has the same error probability as the coset code defined by $\mathbf{A}$ and $\bm v$. Finally, we can find  a non-systematic matrix ${\rm {\bf A}}$ of sufficiently large size  such that $C-\delta \leq R<C-\delta/2$ and ${\rm{FER}}\leq \epsilon$ under the MLD since $E_r(R)>0$ for $R<C$.
\par Next, we prove that the selected non-systematic matrix $\mathbf{A}$ must have full rank. Otherwise,  suppose that the rank of matrix $\mathbf{A}$ is~$r<k$. Suppose that a codeword $\bm c$ is transmitted over a BIOS channel, resulting in $\bm y$. The maximum likelihood~(ML) decoder will find a codeword $\hat{\bm c}$ such that $P(\bm y|\hat {\bm c})$ is maximized and then find an estimated message vector  $\hat{\bm u}\in \mathbb{F}_2^k$ such that $\hat{\bm u}\mathbf{A}=\hat{\bm c}$. Assume that the decoder finds~$j~(j\geq1)$ such codewords. Corresponding to each candidate $\hat{\bm c}$, there are $2^{k-r}$ message vectors $\hat{\bm u}$  such that $\hat{\bm u}\mathbf{A}=\hat{\bm c}$. No matter by what strategy the decoder selects the decoding output from these $j\cdot2^{k-r}$ message vectors, the average decoding error probability  will be~$1-1/(j\cdot 2^{k-r})\geq1/2$ since~$r<k$, which contradicts with~${\rm FER} \leq \epsilon$. The generator matrix  $\mathbf{A}$ of full rank can be  converted, with elementary row transformations and perhaps~(if necessary) some column permutations, into the form of~$[\mathbf{I} ~\mathbf{G}]$, where $\mathbf{I}$ is the identity matrix of order $k$ and $\mathbf{G}$ is a matrix of size $k\times m$. This completes the proof.\hfill $\blacksquare$

\par  \textbf{Remark:} We are saying that the above proof is indirect due to the fact that such a proof does not apply to the case when some constraints are imposed on the matrix $\mathbf{G}$. For example, the above proof cannot answer the question whether or not systematic linear codes with sparse matrices $\mathbf{G}$ of sufficiently large size are capacity-achieving. It is hence instructive to provide a direct proof for more practical codes with constrained matrices $\mathbf{G}$, such as the sparsity measured by the density defined below. 

\begin{definition}
	
	{The density of a matrix $\mathbf{G}$ of size $k\times m$ is defined as 
		\begin{equation}
			S(\mathbf{G})=\frac{{\rm \text{The~number~of~non-zero~elements~in~}}\mathbf{G}}{k\times m}.
	\end{equation}}
\end{definition}
\section{Main Results}\label{sec3}
\subsection{Coding Theorem for  Systematic BGM Codes}

In this paper, we consider the following linear code ensemble~\cite{Ma2016Coding}, which is referred to as systematic Bernoulli generator matrix~(BGM) code ensemble.
\par \emph{Systematic BGM code ensemble:} A systematic BGM code transforms $\bm u\in \mathbb{F}_{2}^{k}$ into $(\bm u,\bm x)$ by $\bm x=\bm u{\rm {\bf G}}\in \mathbb{F}_{2}^{m}$, where ${\rm {\bf G}}$ is a random matrix of size $k \times m$ with each element $G_{i,j}~(0\leq i\leq k-1,~0\leq j \leq m-1)$ being generated independently according to the distribution with a success probability ${\rm Pr}\{G_{i,j}=1\}=\rho \leq 1/2$. Such a matrix is referred to as a Bernoulli  random matrix, which can be  denoted by $\mathbf{G}(\rho)$ for convenience, if the parameter $\rho$ need be emphasized.

\par 
{From the definition above, we see that the average density of the matrices $\mathbf{G}$ in BGM code ensemble is $\rho$.} For $\rho \ll 1/2$, the systematic BGM code ensembles are a class of LDGM codes. It has been proved that, in terms of BER, the systematic BGM code ensembles are capacity-achieving for BIOS memoryless channels~\cite{Ma2016Coding,Cai2020SCLDGM}. However,  with the proof in~\cite{Ma2016Coding,Cai2020SCLDGM}, we cannot conclude that\footnote{Notice that capacity-achieving in terms of BER does not imply capacity-achieving in terms of FER, as illustrated by a counter example presented in~\cite{Cai2020SCLDGM}.} the systematic BGM code ensembles are also capacity-achieving for BIOS memoryless channels in terms of FER. In this paper,  we show that Theorem $1$ also holds even for $\rho\ll1/2$.  In the proof, the systematic bits and the parity-check bits play different roles. Receiving noisy systematic bits provides us a list of the source output, while receiving noisy parity-check bits helps us to select the correct one from the list. To see that the proof is non-trivial and why we need develop new proof techniques, we emphasize the speciality of the BGM code ensemble compared with conventional code ensembles.
\begin{itemize}
	\item The non-zero codewords in the BGM code ensemble are semi-random, meaning that a given message vector $\bm u\in \mathbb{F}_2^k$ is encoded into a codeword of the form $(\bm u, \bm x)$ with a random parity-check vector $\bm x$.
	\item There are $k$ codewords corresponding to the message vectors with Hamming weight one, whose parity-check vectors are independent and identically distributed as a Bernoulli process with the success probability $\rho$.
	\item For $\omega \geq 2$, there are $\binom{k}{\omega}$ codewords corresponding to the message vectors with Hamming weight $\omega$, whose parity-check vectors are sums of $\omega$ Bernoulli processes and hence are identically distributed but may not be (pair-wise) independent.
\end{itemize}
We have the following theorem, which asserts the existence of capacity-achieving sparse generator matrix codes.
\begin{theorem}\rm \label{main_theorem}
	Consider the  systematic BGM code ensemble defined by the generator matrices of the form $[\mathbf{I}~\mathbf{G}]$, where $\mathbf{I}$ is the identity matrix of order $k$ and $\mathbf{G}$ is a sample from the Bernoulli  random matrix of size $k\times m$ { with positive $\rho\leq1/2$}. For three arbitrarily small positive numbers $\epsilon$, $\delta$ and $\eta$, one can always find a sufficiently large integer $k_0$ such that, for all $k \geq k_0$ and $m=\lfloor k(1-C+\delta)/(C-\delta)\rfloor$, there exists a matrix $\mathbf{G}$ of size $k\times m$ {with $S(\mathbf{G})\leq \rho+\eta $} satisfying that  {$m/k\leq (1-C+\delta)/(C-\delta)$}  and the MLD error rate ${\rm FER} \leq \epsilon$.
\end{theorem} 

\subsection{Coding Theorem for Bernoulli Parity-check Codes}
\par \emph{BPC code ensemble: }A BPC code ensemble is defined by  $\mathscr{C}=\{\bm u\in \mathbb{F}_2^k|\bm u\mathbf{G}=\bm 0\}$, where $\mathbf{G}$ is a Bernoulli random matrix of size $k\times m$. The parity-check matrix of  a BPC code is given by $\mathbf{H}=\mathbf{G}^T$, and the code rate is not lower than the design code rate given by $(k-m)/k$. For $\rho\ll1/2$, the BPC code ensemble is a class of LDPC codes.
\par It is not difficult to see that, over BIOS channels, the performance of a parity-check code specified by $\mathbf{H} = \mathbf{G}^T$ is equivalent to the systematic code defined by $[\mathbf{I}~\mathbf{G}]$ with the parity-check vector $\bm X^m = \bm U^k\mathbf{G}$ being transmitted over a noiseless channel. Keeping this equivalence in mind, we see that the following coding theorem is derived  essentially  for the BPC codes over BIOS channels. 

{\textbf{Remark:} The BPC code ensemble is specified by the success probability $\rho$ of the Bernoulli process to generate the parity-check matrices, which is different from the conventional LDPC code ensemble defined with the degree distributions. For example, the 8 LDPC ensembles in~\cite{Litsyn2003} are specified with fixed~(average) column weights and column weights in the parity-check matrices. The density of BPC is $S(\mathbf{H})=\rho>0$ while the density in~\cite{Litsyn2003} is $S(\mathbf{H})\rightarrow 0$ as the code length tends to infinity. In addition, the members in the BPC code ensemble are non-equiprobable while the members in conventional LDPC code ensemble are equiprobable. }
\begin{theorem}\rm \label{LDPC_theorem}
	Consider the BGM code defined by the  matrices of the form $[\mathbf{I}~\mathbf{G}]$, where  $\mathbf{G}$ is a sample from the Bernoulli  random matrix of size $k\times m$ { with positive $\rho\leq1/2$}. Suppose that the message vector $\bm U^k$ is transmitted over a BIOS channel with a capacity of $C$ and that the parity-check vector $\bm X^m = \bm U^k \mathbf{G}$ is transmitted over a noiseless channel. For three arbitrarily small positive numbers $\epsilon$, $\delta$ and $\eta$, one can always find a sufficiently large integer $k_0$ such that, for all $k \geq k_0$ and $m=\lfloor k(1-C+\delta)\rfloor$, there exists a matrix $\mathbf{G}$ of size $k\times m$ with {$\rho(\mathbf{G})<\rho+\eta$} such that the design code rate $R = (k-m)/k \geq C-\delta$  and the MLD error rate ${\rm FER} \leq \epsilon$.
\end{theorem} 

\subsection{Performance of Finite Length BGM Codes}
\par  We have proved in Theorem~\ref{main_theorem} that, like the totally random linear codes,  systematic BGM codes can achieve the capacity of the   BIOS memoryless channels.  Now we provide tools to analyze the performance of  systematic BGM codes in the finite length region. For simplicity, we focus on BPSK signaling over AWGN channels. Consider the BPSK mapping $\phi:\mathbb{F}_2^{k+m}\rightarrow\{+1,-1\}^{k+m}$ taking a codeword $\bm c$ to a bipolar signal~(also referred to as a codeword for convenience) $\bm s=\phi(\bm c)$ by $s_t=1-2c_t$ for $0\leq t\leq t-1$. Particularly, we denote $\bm s_0=\phi(\bm 0)$. With this mapping, the Euclidean distance between two codewords $||\bm s_i-\bm s_j||=2\sqrt{d}$, where $d$ is the Hamming weight distance between the two corresponding binary codewords $\bm c_i$ and $\bm c_j$. We use $\overrightarrow{\bm s_i\bm s_j}$ to denote the vector in $\mathbb{R}^{k+m}$ directed from $\bm s_i$ to $\bm s_j$.
\par Deriving lower bounds is important to  justify a suboptimal decoding algorithm. If a decoding algorithm  performs close to the lower bound, on the one hand, we can conclude that the decoding algorithm is near optimal; on the other hand, we can conclude that the bound is tight. We first present a lower bound on the BER of  systematic BGM codes.
\begin{theorem}\rm\label{theorem_BER_bound}
	For the BPSK-AWGN channel, the BER of a systematic BGM code with  matrix $[\mathbf{I}~\mathbf{G}]$ of size $k\times (k+m)$ can be lower bounded by
	\begin{equation}
		\label{eq_BER}
		\begin{aligned}
			{\rm BER}\geq \frac{1}{k}\sum\limits_{i=0}^{k}Q\Big(\frac{\sqrt{\omega_i}}{\sigma}\Big),
		\end{aligned}
	\end{equation}
	where $\omega_i$ is the Hamming weight of the $i$-th row of $[\mathbf{I}~\mathbf{G}]$, $\sigma^2$ is the variance of the noise and $Q(x)$ is the tail probability that the normalized Gaussian random variable takes a value not less than $x$.
\end{theorem}
\par To derive the lower bounds on FER for a systematic BGM code over the BPSK-AWGN channel, without loss of generality, assume that the all zero codeword $\bm c=\bm 0\in \mathbb{F}_2^{k+m}$ (actually $\bm s_0=1-2\bm c$) is transmitted over an AWGN channel, resulting in $\bm y\in \mathbb{R}^{k+m}$. 
The MLD delivers the transmitted codeword if and only if $\bm y\in \Omega$, the Voronoi region of $\bm s_0$, which can be formed by intersecting $2^{k+m}-1$ half-spaces around $\bm s_0$~\cite{1996Agrell}.  To be precise, $\Omega=\cap_{i=1}^{2^{k+m}-1}\Omega_i$ and $\Omega_i=\{\bm y\in \mathbb{R}^{k+m}\big| \Vert\bm y-\bm s_0\Vert\leq \Vert{\bm y-\bm s_i}\Vert\}$ is the half-space defined by the perpendicular bisection plane of $\overrightarrow{\bm s_0\bm s_i}$.
\par Intuitively, the probability ${\rm Pr}\{ \bm y\in \Omega|\bm s_0\}$ of correct decoding can be upper-bounded by ${\rm Pr}\{ \bm y\in \tilde{\Omega}|\bm s_0\}$ for any enlarged $\tilde{\Omega}\supseteq\Omega$.  In particular, we may derive an upper bound  by forming a list $\mathscr{L} = \{\bm s_1, \bm s_2, \cdots, \bm s_L\}$ and setting $\tilde{ \Omega} = \cap_{i=1}^L\Omega_i$.  To derive a tight bound, we form the list by  a greedy algorithm starting from $\mathscr{L} = \{\bm s_1\}$, where the associated binary vector $\bm c_1$ is one of the lightest rows of $[\mathbf{I}~\mathbf{G}]$. At the $i$-th step for $i > 1$, the greedy algorithm chooses from the remaining rows of $[\mathbf{I}~ \mathbf{G}]$  a row vector, denoted by $\bm c_i$, such that $W_H(\bm c_i)$ is as small as possible and the associated vector $\overrightarrow{\bm s_0\bm s_i}$  is orthogonal  to all $\overrightarrow{\bm s_0\bm s_j}$ with $j < i$. Obviously, this greedy algorithm must terminate with some $L \leq k$. We have the following theorem.
\begin{theorem}\label{theorem_FER_bound} 
	Let $\mathscr{L} =\{\bm s_1,\bm s_2,\cdots, \bm s_L\}$ be a list of codewords such that, for $1\leq i < j \leq L$, $\overrightarrow{\bm s_0\bm s_i}$ and $\overrightarrow{\bm s_0\bm s_j}$ are orthogonal. For the BPSK-AWGN channel, the FER of a systematic BGM code  can be lower bounded by
	\begin{equation}
		\label{eq_FER}
		{\rm FER}\geq 1-\prod_{i=1}^{L}\Big[1-Q\Big(\frac{\sqrt{\omega_i}}{\sigma}\Big)\Big]
	\end{equation}
	where $\omega_i$ is the Hamming weight of the  codeword $\bm c_i$ associated with $\bm s_i \in \mathscr{L}$, $\sigma^2$ is the variance of the noise and $Q(x)$ is the tail probability that the normalized Gaussian random variable takes a value not less than $x$.
\end{theorem}

\section{Proof of the Coding Theorems}
\label{sec4}
\subsection{Partial Mutual Information}
Let $P(1)=p$ and $P(0)=1-p$~be an input distribution of a BIOS memoryless channel. The mutual information between the input and the output is given by

\begin{equation}
	I(p)=(1-p)I_0(p)+pI_1(p)\text{,}
\end{equation}
where
\begin{equation}
	I_0(p)=\sum\limits_{y\in\mathcal{Y}} P(y|0)\log\frac{P(y|0)}{P(y)}\text{,}
	\label{I_{0}}
\end{equation}
\begin{equation}
	I_1(p)=\sum\limits_{y\in\mathcal{Y}} P(y|1)\log\frac{P(y|1)}{P(y)}\text{,}
\end{equation}
and $P(y)=(1-p)P(y|0)+pP(y|1)$. We define $I_0(p)$ $\big({\rm or}~I_1(p)\big)$  as \emph{partial mutual information}. For a BIOS memoryless channel, we have ${\rm max}_{0\leq p \leq 1}I(p)=I(1/2)=I_0(1/2)=I_1(1/2)$, which is the channel capacity. Notice that $I_0(p) > 0$ for $0<p<1$ as long as ${\rm Pr}\{y |P(y|0) \neq P(y|1)\} > 0$. This is a natural assumption in this paper.

\begin{lemma}
	\label{lemma_for_I}	
	\rm 
	The partial mutual information $I_0(p)$ is continuous, differentiable and strictly increasing from $I_0(0)=0$ to the capacity $I_0(1/2)$.
\end{lemma}	

\begin{Proof}
	It can be easily seen that the partial mutual information is continuous and differentiable for $0\leq p\leq 1/2$. By carrying out the differentiation, we can verify that partial mutual information is strictly increasing from $I_0(0)=0$ to the capacity $I_0(1/2)$. 
\end{Proof}	 
	\subsection{Partial Error Exponent}
\begin{lemma}
	\label{lemma_for_rho}\rm 
	For the  code ensemble defined by the generator matrix $[\mathbf{I}~\mathbf{G}(\rho)]$ {with positive $\rho\leq1/2$}, the parity-check vector corresponding to a message vector with weight $\omega$ is a Bernoulli sequence with success probability
\begin{equation}
	\rho_{\omega}\triangleq{\rm{Pr}}\{X_j=1|W_H(\bm U)=\omega\}=\frac{1-(1-2\rho)^{\omega}}{2}\text{,}
\end{equation}
where $W_H(\cdot)$  is the Hamming weight function. Furthermore, for any given positive integer $T\leq k$,
\begin{equation}
	\begin{aligned}
		P(\bm x|\bm u)&\triangleq {\rm Pr}\{\bm X=\bm x|\bm U=\bm u \}\\
		&\leq P(\bm 0|\bm u)\leq(1-\rho_{T})^{m}\text{,}
	\end{aligned}
\end{equation}
for all $\bm u\in \mathbb{F}_{2}^{k}$ with $W_H(\bm u)\geq T$ and $\bm x\in \mathbb{F}_{2}^{m}$.
\end{lemma}	
\begin{Proof}
See Appendix.
\end{Proof}

\par\textbf{Remark:} Lemma~\ref{lemma_for_rho} states that, for a message vector $\bm u$ with high weight, the corresponding parity check vector is convergent in distribution to a Bernoulli process with success probability $1/2$, since $\rho_{\omega}\rightarrow 1/2$ as $\omega\rightarrow \infty$.
\par The proof of coding theorem in~\cite[Theorem 5.6.1]{Gallager1968} with error exponent is so general that it can apply to many channels~(memory/memoryless and discrete/non-discrete). To achieve the generality, the code ensemble generated in the proof should satisfy the following two constraints:
\begin{enumerate}
	\item[1)] The codewords should be selected following some identical distribution.
	\item[2)] The codewords should be selected having pair-wise independence.
\end{enumerate}
Intuitively, these constraints can be relaxed when Theorem 5.6.1 in~\cite{Gallager1968} is specialized to the case of BIOS channels, Bernoulli sources and binary linear codes. 
In this paper, we derive the partial error exponent for BIOS channels by assuming that the codeword $\bm 0$ is transmitted. The derivation suggests that  the pair-wise independence is not required.
\begin{theorem}
	\label{partial_error_exponent}\rm 
	Suppose that the codeword $\bm 0\in\mathbb{F}_2^{n}$ is transmitted over a BIOS channel. Let $\mathscr{L}=\{\bm x_1,\bm x_2,\cdots,\bm x_L\}$ be a random list, where $\bm x_i\in\mathbb{F}_2^n$ is a segment of a Bernoulli process with success probability $p$. Then the probability that there exists some $i$ such that $\bm x_i$ is more likely than $\bm 0$, denoted by ${\rm Pr}\{{\rm error}|\bm 0\}$, can be upper bounded by

	\begin{equation}
		\label{conditional_pro}
		{\rm Pr}\{{\rm error}|\bm 0\}\leq \exp\Big[-nE(p,R)\Big]\text{,}
	\end{equation}
	where
	\begin{equation}
		R=\frac{1}{n}\log L,
	\end{equation}
	\begin{equation}
		E(p,R)=\max\limits_{0\leq\gamma\leq 1}(E_0(p,\gamma)-\gamma R)\text{,}
	\end{equation}
	and
	\begin{equation}
		\begin{aligned}
			E_0(p,\gamma)=-\log\Bigg\{\sum\limits_{ y\in \mathcal{Y}}(P( y| 0))&^{\frac{1}{1+\gamma}}\Big[(1-p)(P( y| 0))^{\frac{1}{1+\gamma}}+p(P( y| 1))^{\frac{1}{1+\gamma}}\Big]^\gamma\Bigg\}\text{.}
		\end{aligned}	
		\label{E independent}
	\end{equation}
	Furthermore, $E(p,R)>0$ if $0<R<I_0(p)$.
\end{theorem}

\begin{Proof}
	Denote by $A_i$ the event that $\bm x_i$ is more likely than $\bm 0$ given a received sequence $\bm y$. For the decoding error, we have
	\begin{equation}
		\begin{aligned}
			{\rm Pr}\{{\rm error}|\bm 0\}&=\sum\limits_{\bm y\in \mathcal{Y}^{n}}P(\bm y|\bm 0)\cdot {\rm {Pr}}\Big\{{\bigcup\limits_{i=1}^L A_i\Big\}}\\
			&\overset{(*)}{\leq}  L^\gamma\sum\limits_{\bm y\in \mathcal{Y}^{n}}P(\bm y|\bm 0)\bigg({\rm {Pr}}\{P(\bm y|\bm x)\geq P(\bm y|\bm 0)\}\bigg)^\gamma\text{,}
		\end{aligned}
		\label{P}	
	\end{equation}
	for any given $0\leq \gamma\leq 1$, where the inequality $(*)$ follows from~\cite[Lemma in Chapter 5.6]{Gallager1968}. From Markov inequality, for  $s=1/(1+\gamma)$ and a given received vector $\bm y$, the probability of a vector $\bm x$ being more likely than $\bm 0$ is upper bounded by
	\begin{equation}
		\begin{aligned}
			{\rm Pr}\{P(\bm y|\bm x)\geq P(\bm y|\bm 0)\}&\leq  \frac{{  \rm{\bf E}}[(P(\bm y|\bm x))^{s} ]}{(P(\bm y|\bm 0))^{s}}\\
			&=\sum\limits_{\bm  x} P(\bm x)\frac{(P(\bm y|\bm x))^{s}}{(P(\bm y|\bm 0))^{s}}\text{.}
		\end{aligned}
		\label{Markov}
	\end{equation}
	
	Substituting this bound into \eqref{P}, we have
	\begin{equation}
		\begin{aligned}
			&{\rm Pr}\{{\rm error}|\bm 0\}\leq L^\gamma \sum\limits_{\bm y\in \mathcal{Y}^{n}}P(\bm y|\bm 0) \Big[\sum\limits_{\bm  x} P(\bm x)\frac{(P(\bm y|\bm x))^{s}}{(P(\bm y|\bm 0))^{s}}\Big]^\gamma\\
			&\overset{(*)}{=}L^\gamma\prod\limits_{i=0}^{n-1}\Bigg\{\sum\limits_{ y_i\in \mathcal{Y}}(P(y_i|0))^{1-s\gamma}\Big[\sum\limits_{x_i\in \mathbb{F}_2} P(x_i)(P(y_i|x_i))^{s}\Big]^\gamma\Bigg\}\\
			&\overset{(**)}{\leq}\exp\Big[-nE(p,R)\Big]\text{,}
		\end{aligned}
	\end{equation}
	where the equality $(*)$ follows from the memoryless channel assumption and the inequality $(**)$ follows by recalling  that $s=1/(1+\gamma)$ and denoting
	\begin{equation}
		E(p,R)=\max\limits_{0\leq\gamma\leq 1}(E_0(p,\gamma)-\gamma R)\text{,}
	\end{equation}
	and
	\begin{equation}
		\begin{aligned}
			E_0(p,\gamma)=-\log\Bigg\{\sum\limits_{ y\in \mathcal{Y}}(P( y| 0))^{\frac{1}{1+\gamma}}&\Big[(1-p)(P( y| 0))^{\frac{1}{1+\gamma}}+p(P( y| 1))^{\frac{1}{1+\gamma}}\Big]^\gamma\Bigg\} \text{.}
		\end{aligned}	
	\end{equation}
	
	Considering $E_0(p,\gamma)-\gamma R$ for a given $p$, we have $E_0(p,0)-0\cdot R=0$ and
	\begin{equation}
		\frac{\partial E_0(p,\gamma)}{\partial \gamma}-R\Bigg|_{\gamma=0}=I_{0}(p)-R\text{.}
	\end{equation}
	Hence, $E(p, R) > 0 $ if $ R < I_0(p)$.
\end{Proof}

\par \textbf{Remarks:} For the partial error exponent, we have the following remarks:
\begin{itemize}
	\item {From the above proof, we see that the members in the list need to have the same distribution
		or have the same upper bounds of probability but the condition of pair-wise independence
		is not necessary, which is distinguished from the proof in~\cite[Chapter 5]{Gallager1968}. In particular, the
		random list $\mathscr{L}$ assumed in Theorem~\ref{partial_error_exponent} has identically distributed members, each of which is a segment of a Bernoulli process with success probability $p$. But the members in $\mathscr{L}$ can be correlated. Even in the extreme case when the members in $\mathscr{L}$ are strongly correlated, say $\bm x_i = \bm x_1$ for all $i >1$, the bound for ${\rm Pr}\{ {\rm error}| \bm 0 \}$ in~\eqref{conditional_pro} still holds.}
	\item The partial error exponent derived in this paper applies only to BIOS channels while the error exponent in~\cite[Chapter 5]{Gallager1968} can apply to any stationary and memoryless channel.
	
\end{itemize}


\subsection{List Coset Decoding}
\label{AchievableForBGMCs}
\par To prove the coding theorem, we need a list decoding as described in the following theorem. List decoding was first introduced by Elias to explore average error probability of block codes in BSC~\cite{elias1957list} and  was  used  to derive average error probability bounds for general discrete memoryless  channels~\cite{shannon1967lower}. Practical list decoding algorithms have been developed correspondingly for decoding, say, convolutional codes~\cite{seshadri1994list}, algebraic codes~\cite{1999Guruswami} and polar codes~\cite{2012CRC,tal2015list}. Different from the commonly-accepted list decoding, which attempts to find  a list of $L$ most probable candidate codewords, our presented list decoding aims at forming a list~(with a constrained size) to contain the transmitted~(uncoded) vector with high probability. Some members in the list $\mathscr{L}$ delivered by our list decoding may not be the $L$ most probable candidates. 
\begin{theorem}\rm 
	\label{theorem_for_SBGMC}
	Let   $\epsilon$ and $\delta$ be two arbitrarily small positive numbers. Suppose that  $\bm u\in\mathbb{F}_2^k$  is transmitted over a BIOS channel, resulting in $\bm v\in \mathcal{V}^k$. Then there exists a list decoding algorithm to deliver a list $\mathscr{L}$ of size $L\leq\exp[k(1-C+\delta)]$ such that  ${\rm Pr}\{\bm u \notin \mathscr{L}\}\leq \epsilon$ for sufficiently large $k$.
\end{theorem}

\begin{Proof}
Generate a totally random binary  matrix $\mathbf{A}$ of size $k\times \widetilde{m}$ with $\widetilde{m}=\lfloor k(1-C+\delta)\rfloor$. Given the received vector $\bm v\in \mathcal{V}^k$,  we perform the following list decoding algorithm. 
	\par \emph{List coset decoding algorithm~(LCDA):} For each $\bm z\in \mathbb{F}_2^{\widetilde{m}}$, find a vector $u(\bm z) \in \mathbb{F}_2^k$ from the coset code $\mathscr{C}(\bm z) = \{\bm x\in \mathbb{F}_2^k|\bm x\mathbf{A}=\bm z\}$ such that $P(\bm v| u(\bm z)) \geq P( \bm v | \bm w) $ for all $\bm w \in\mathscr{C}(\bm z)$. The list is then given by $ \mathscr{L} = \{u(\bm z) | \bm z \in \mathbb{F}_2^{\widetilde{m}} \}$.  Obviously, the list size $L \leq \exp[k(1-C+\delta)]$. Next we show that ${\rm Pr}\{\bm u\notin \mathscr{L}\}\leq \epsilon$ for sufficiently large $k$.
	\par Notice that transmitted vector $\bm u$ is not in $\mathscr{L}$ if and only if the MLD of the coset code defined by $\mathbf{A}$ and $  \bm z= \bm u \mathbf{A} $ is in error. Without loss of generality, we assume that $\bm u=\bm 0^k$~(for $\bm u\neq \bm 0^k$, we consider the coset code instead). 
	Due to the randomness of the matrix $\mathbf{A}$, $\mathscr{C}(\bm 0)=\{\bm 0, \bm x_1, \bm x_2, \cdots, \bm x_{M-1}\}$ is a  random codebook of size $M = \exp(k-{\widetilde{m}})$ with each nonzero codeword $\bm x_i$ being a segment of a  Bernoulli process with success probability $1/2$. From Theorem~\ref{partial_error_exponent}, we have~(despite of the dependence between $\bm x_i$ and $\bm x_j$)
	
	\begin{equation}
		{\rm Pr}\{u(\bm 0)\neq \bm 0\}\leq \exp\Big[-kE\Big(\frac{1}{2},R_M\Big)\Big]\text{,}
	\end{equation}
	where
	\begin{equation}
		R_M=\frac{k-{\widetilde{m}}}{k}\text{.}
	\end{equation}
	Since $R_M<C-\delta/2$ for sufficiently large $k$, we can conclude that the error probability ${\rm Pr}\{ u(\bm 0)\neq \bm 0\}$ goes to zero exponentially with increasing $k$. Hence, we have ${\rm Pr}\{\bm u \notin \mathscr{L}\}$ goes to zero exponentially as $k\rightarrow \infty$.
\end{Proof}

\par \textbf{Remark:}  Theorem~\ref{theorem_for_SBGMC} is intuitively correct since for a given $\bm v$, we may form the list $ \mathscr{L}$ by listing all  vectors $\bm u$ which are jointly typical with $\bm v$. From~\cite[Section 15.2.1]{Cover2006} and ~\cite[Section 15.2.2]{Cover2006}, we can conclude that~${\rm Pr}\{\bm u \notin \mathscr{L}\}\leq \epsilon$ and $L\leq\exp[k(H(U|V)+\delta)]$~(noticing that $H(U|V)=1-C$). However, we cannot see whether or not ${\rm Pr}\{\bm u \notin \mathscr{L}\}$ approaches zero exponentially  as $k\rightarrow \infty$.  To see the difference, we need point out that some members in $\mathscr{L}$ delivered by the LCDA may not be jointly typical with the received sequence.  

\subsection{Proof of Theorem~\ref{main_theorem}}

\textit{Proof of Theorem~\ref{main_theorem}:}
	 For {$m/k\leq (1-C+\delta)/(C-\delta)$}, we have the code rate $R=k/(k+m)\geq C-\delta$. For BIOS memoryless channels, without loss of generality, suppose that $(\bm 0,\bm0{\rm {\bf G}})$ is transmitted. Let $\epsilon>0$ be an arbitrarily small number. Upon receiving $(\bm v^k,\bm y^m)$, we use the following two-step decoding. First, list all sequences $\bm {\tilde u}$ using LCDA with $\bm v^k$. Second, find from the list a sequence $\hat{\bm u}$ such that $P(\bm y|\hat{\bm u}{\rm {\bf G}})$ is maximized.
	\par There are two types of errors. One is the case when $\bm 0$ is not in the list. This type of errors, from Theorem~\ref{theorem_for_SBGMC}, can have arbitrarily small probability as long as $k$ is sufficiently large.
		\par The other case is that $\bm 0$ is in the list but is not the most likely one.  In this case, denote the list as ${\mathscr{{L}}}=\{{ {\bm u}}_0=\bm 0,{{\bm u}}_1,\cdots,{ {\bm u}}_L\}$.  From Theorem~\ref{theorem_for_SBGMC}, we have $L\leq \exp\big[k(1-C+\delta)\big]$. Given the received sequence $\bm y$ and the list  ${\mathscr{{L}}}$, the decoding output~$\hat{\bm U}$ is a random sequence over the code ensemble due to the randomness of $\mathbf{G}$.  Given a received sequence $\bm y$, denote by~$A_{i}$ the event that ${\bm u}_i\mathbf{G}$ is more likely than $\bm 0$.  We have
	\begin{equation}
		\label{error0}
		\begin{aligned}
			{\rm{Pr}}\{{\rm error}|\bm u^{k}\}
			&\leq \exp\Big\{-kE\Big(\frac{1}{2},R_M\Big)\Big\}+\sum\limits_{\bm y\in \mathcal{Y}^{m}}P(\bm y|\bm 0)\cdot {\rm {Pr}}\Big\{{\bigcup\limits_{i=1}^LA_i\Big\}}\text{.}
		\end{aligned}	
	\end{equation}
	\par  We partion the list $\mathscr{L}$ according to the weight of  ${\bm u}_i~(0\leq i\leq L)$ and denote by $\mathscr{L}_\omega$ all the sequences of $\bm u_i\in\mathscr{L}$ with $W_H(\bm u_i)=\omega$. Thus, we have
	\begin{equation}
		\mathscr{L}=\bigcup\limits_{\omega=0}^{k}\mathscr{L}_{\omega}\text{,}
	\end{equation}
	and
	\begin{equation}
		\big|\mathscr{L}_{\omega}\big|\leq\binom{k}{\omega}\text{.}
		\label{boundnumber}
	\end{equation}
	\par For any positive integer  $T<k$, the error event can be split into two sub-events depending on whether $W_H(\bm U)\geq T$ or not. Thus, we have
	
	\begin{equation}\small
		\begin{aligned}
			\sum\limits_{\bm y\in \mathcal{Y}^{m}}P(\bm y|\bm 0){\rm {Pr}}\Big\{{\bigcup\limits_{i=1}^LA_i\Big\}}
			&\leq\sum\limits_{\bm y\in \mathcal{Y}^{m}}P(\bm y|\bm 0)\sum\limits_{w=1}^{T-1}{\rm Pr}\left\{\bigcup\limits_{\bm u\in \mathscr{L}_w}P(\bm y|\bm u{\rm {\bf G}})\geq P(\bm y|\bm 0)\right\}\\
			&+\sum\limits_{\bm y\in \mathcal{Y}^{m}}P(\bm y|\bm 0){\rm Pr}\left\{\bigcup\limits_{\substack{w\geq T \\ \bm u\in\mathscr{L}_w}}P(\bm y|\bm u{\mathbf {G}})\geq P(\bm y|\bm 0)\right\}\text{,}
		\end{aligned}
		\label{bound1}
	\end{equation}
	for any $ 0\leq\gamma\leq 1$.
	
	For $\omega \geq1$, we define FER($\omega$) as
	\begin{equation}
		\begin{aligned}
			{\rm FER}(\omega)&=\sum\limits_{\bm y\in \mathcal{Y}^{m}}P(\bm y|\bm 0){\rm Pr}\left\{\bigcup\limits_{\bm u\in \mathscr{L}_w}P(\bm y|\bm u{\rm {\bf G}})\geq P(\bm y|\bm 0)\right\}\text{.}
		\end{aligned}
	\end{equation}
	
	We know from Lemma~\ref{lemma_for_rho} that, for any given $\bm u$ with $W_H(\bm u) = \omega$, the parity-check vector $\bm x$ is a segment of Bernoulli process with success probability $\rho_{\omega}$.
	Hence, the conditional probability mass function $P(\bm x |\bm u)$ for $\bm u \in \mathscr{L}_{\omega}$ is the same, denoted by $P_{\omega}(\bm x)$. We have
	\begin{equation}
		\begin{aligned}
			{\rm FER}(\omega) &\overset{(*)}{\leq} \sum\limits_{\bm y\in \mathcal{Y}^{m}}P(\bm y|\bm 0)\Bigg[\binom{k}{\omega}\sum_{\bm x\in\mathbb{F}_{2}^{m}} P_{\omega}(\bm x)\frac{(P(\bm y|\bm x))^{s}}{(P(\bm y|\bm 0))^{s}}\Bigg]^\gamma\\	
			&\overset{(**)}{\leq}\exp\Big[-mE(\rho_\omega,R_\omega)\Big]\text{,}	
		\end{aligned}
		\label{bound2}
	\end{equation}
	where the inequality $(*)$ follows  from the proof of Theorem~\ref{partial_error_exponent} and \eqref{boundnumber}, and inequality $(**)$ follows from the proof of Theorem~\ref{partial_error_exponent} and by denoting
	\begin{equation}
		R_\omega=\frac{1}{m}\log{\binom{k}{\omega}}.
	\end{equation}

	\par For $W_H(\bm u)\geq T$, we have
	\begin{equation}
		\begin{aligned}
			{\rm Pr}&\{P(\bm y|\bm u{\rm {\bf G}})\geq P(\bm y|\bm 0)\}\overset{(*)}{\leq}\frac{{\rm{\bf E}}[(P(\bm y|\bm u {\rm {\bf G}}))^s]}{(P(\bm y|\bm 0))^s}\\
			&=\sum\limits_{ \bm x\in \mathbb{F}_2^m}P(\bm x|\bm u)\frac{(P(\bm y|\bm u {\rm {\bf G}}))^s}{(P(\bm y|\bm 0))^s}\\
			&\overset{(**)}{\leq}\Bigg[\frac{1+(1-2\rho)^T}{2}\Bigg]^m\sum\limits_{ \bm x\in \mathbb{F}_2^m}\frac{(P(\bm y|\bm x))^s}{(P(\bm y|\bm 0))^s},
		\end{aligned}
	\end{equation}
	where the inequality~$(**)$ follows from the Markov inequality for any $s>0$,  and the inequality $(**)$ follows from Lemma~\ref{lemma_for_rho}. Thus,  we have
	
	\begin{equation}
		\begin{aligned}	
			&\sum\limits_{\bm y\in \mathcal{Y}^m}P(\bm y|\bm 0){\rm Pr}\left\{\bigcup\limits_{\substack{w\geq T \\ \bm u\in\mathscr{L}_w}}P(\bm y|\bm u{\mathbf {G}})\geq P(\bm y|\bm 0)\right\}\\	
			&\leq \sum\limits_{\bm y\in \mathcal{Y}^m}P(\bm y|\bm 0)\Bigg\{\sum\limits_{\omega\geq T}|\mathscr{L}_{\omega}|
			\cdot\Bigg[\frac{1+(1-2\rho)^T}{2}\Bigg]^{m}\sum\limits_{\bm x \in \mathbb{F}_2^m} \frac{(P(\bm y|\bm x))^{s}}{(P(\bm y|\bm 0))^{s}}\Bigg\}^{\gamma}\\
			&\overset{(*)}{\leq}\exp\Big[{k\gamma (1-C+\delta)}\Big]\Big[1+(1-2\rho)^T\Big]^{m\gamma}\cdot \Bigg[\sum\limits_{ y_i\in \mathcal{Y}}(P(y_i|0))^{1-s\gamma}\Bigg(\sum\limits_{ x_i\in \mathbb{F}_2}\frac{1}{2}\cdot{(P( y_{i}|x_{i}))^{s}}\Bigg)^{\gamma}\Bigg]^{m}\\
			&\overset{(**)}{\leq} \exp\Big[{-mE\Big(\frac{1}{2},{R}_T\Big)}\Big]\text{,}
		\end{aligned}
		\label{latterbound}
	\end{equation}
	where the inequality $(*)$ follows from Theorem~\ref{theorem_for_SBGMC} and the  BIOS memoryless channel assumption, and the inequality $(**)$ follows from the proof of Theorem~\ref{partial_error_exponent} and by denoting
	
	\begin{equation}
		{R}_T= \log[1+(1-2\rho)^T]+ \frac{k}{m} (1-C+\delta)\text{.}
	\end{equation}

	\par Thus, we have
	\begin{equation}
		\begin{aligned}
			{\rm{Pr}}\{{\rm error}|\bm u^k\}&\leq \exp\Big[-kE\Big(\frac{1}{2},R_M\Big)\Big]+\sum\limits_{\omega=1}^{T-1}\exp\Big[-mE(\rho
			_{\omega},R_\omega)\Big]+\exp\Big[{-mE\Big(\frac{1}{2},{R}_T\Big)}\Big]\text{.}
			\label{result}
		\end{aligned}
	\end{equation}
	From Theorem~\ref{theorem_for_SBGMC}, we  see that the first term can be made not greater than $\epsilon/3$  for sufficiently large $k$. Now letting $k \rightarrow \infty$ and $T \rightarrow \infty$, we have $ {R}_T < C-\delta/2 $ since $\rho \leq 1/2$. We have from Theorem~\ref{partial_error_exponent} that $E(1/2, {R}_T) > 0 $ and the third term in the right hand side~(RHS) of the inequality~\eqref{result} can be made not greater than $\epsilon/3$ for sufficiently large $k$ since $m$ increases linearly with $k$.
	By fixing $T$ and for $\omega < T$,  $\log{\binom{k}{\omega}}$ increases only logarithmically with $k$, we have $R_\omega \rightarrow 0 $ as $k \rightarrow \infty$.
	Since $I_0(\rho_\omega)> 0$, we have $E(\rho_{\omega},R_\omega) > 0$ for sufficiently large $k$, implying that the second term in the RHS of~\eqref{result} can also be made not greater than~$\epsilon/3$.
	Now we have
	\begin{equation}
		{\rm Pr}\{{\rm error}|\bm 0\} \leq \epsilon.
	\end{equation}
	Therefore, we have the error probability in the ensemble 
	\begin{equation}
		{\rm Pr}\{{\rm error} \} = \sum_{\bm u^k \in \mathbb{F}_2^{k}} 2^{-k} {\rm Pr}\{{\rm error}|\bm u^k\} \leq  \epsilon. 
	\end{equation}
	
	\par { From the weak law of large numbers, we have, for sufficiently large $k$
\begin{equation}
		{\rm Pr}\{|S(\mathbf{G})-\rho|\leq\eta\}\geq 1-\eta.
\end{equation}
Then we assert that there must exist a matrix $\mathbf{G}$ with $S(\mathbf{G}) \leq \rho+ \eta$ and ${\rm Pr}\{{\rm error}|\mathbf{G}\} \leq \epsilon/(1-\eta)$.  Otherwise, we will have ${\rm Pr}\{{\rm error}\} >(1-\eta) \epsilon/(1-\eta) =\epsilon$.
}
{Thus, we complete the proof of Theorem~\ref{main_theorem}.}

\par \textbf{Remarks:} The proof of Theorem~\ref{main_theorem} provides us new insights into the capacity-achieving code ensembles over BIOS channels.
\begin{itemize}
	\item The codewords can be semi-random in the sense that a codeword is partitioned into two parts: the deterministic message bits and the randomly generated parity-check bits. From the two-step decoding algorithm, we see that the systematic bits and the parity-check bits can play different roles. The noisy systematic bits are exploited to form a list and the noisy parity-check bits are exploited to select  a candidate from the list. The list is formed to contain the transmitted~(uncoded) with sufficiently high probability but has a constrained size so that its equivocation can be recovered by the parity-check bits. Such a proof can be generalized to the scenarios  where the systematic bits and the parity-check bits are transmitted through different BIOS channels, as pointed out in~\cite{Wang2022}.      
	\item For BIOS memoryless channels, the pair-wise independence for the codewords is not necessary. We believe  that this can be generalized to geometrically uniform codes~\cite{Forney1991} over AWGN channels.  
	\item The codewords need not to be uniformly distributed. Actually, the codewords need not have identical distributions. The weight of the parity-check vector generated by the light message vector is also light. The mutual information induced by the light~(sparse) parity-check vectors is less but enough to distinguish the transmitted one from its neighbours~(sparse errors).
		\item The generator matrix~(corresponding to the parity-check part) is not necessarily uniformly distributed. Actually, its elements can be independent and identically distributed~(i.i.d) with a success probability $\rho\ll 1/2$. This admits the proof that sparse codes are capacity-achieving over BIOS
		channels, for which case the proof in~\cite{Gallager1968} is not applicable.
		\item The codewords are not necessarily identically distributed. Actually, the denser the message bits are, the more uniform the parity-check bits are.

\end{itemize}

\subsection{Proof of Theorem~\ref{LDPC_theorem}}

\par  To prove that the BPC codes are capacity-achieving, we make an assumption that the message vector   $\bm U^{k}$ is  transmitted through (possibly) noisy BIOS~$1$ and the parity-check vector $\bm X^m$ is transmitted through noiseless channel BIOS~$2$, resulting in $\bm V^{k}$ and $\bm Y^m~(=\bm X^m)$, respectively.

\par \emph{Proof of Theorem~\ref{LDPC_theorem}:}
	Let $\epsilon>0$ be an arbitrarily small number. Upon receiving $(\bm v^{k},\bm y^m)$, we consider the  two-step decoding. First, list all sequences $\bm {\tilde u}$ using LCDA with $\bm v^{k}$ and we have the list size $L\leq \exp[k(1-C+\delta/2)]$. Second, find from the list a sequence $\hat{\bm u}$ such that $\hat{\bm u}\mathbf{G}=\bm y$. If there are two or more sequences  satisfying the equation, we simply choose at random one sequence as the decoding result. 
	\par  There are two types of errors. One is the case when $\bm u^{k}$ is not in the list. From Theorem~\ref{theorem_for_SBGMC}, we have that the error probability is less than $\exp[-kE(\frac{1}{2},R_M)]$, where $R_M=(k-m)/k$.
	\par The other case is that $\bm u^{k}$ is in the list but there exists another sequence $\tilde{\bm u}$ such that $\tilde{\bm u}\mathbf{G}=\bm y$. In this case, denote the list as ${\mathscr{L}}=\{{\bm u}_0=\bm u,{\bm u}_1,{\bm u}_2,\cdots,{\bm u}_L\}$ and we have $L\leq \exp\big[k(1-C+\delta/2)\big]$. Given the received sequence $\bm y$ and  the list  ${\mathscr{L}}$, the decoding output $\hat{\bm U}$ is a random sequence over the code ensemble due to  the randomness of $\mathbf{G}$. Denote by $A_{ i}$ the event that $\bm u_i\mathbf{G}=\bm y$, which is essentially the same as the event that $(\bm u_i-\bm u)\mathbf{G}=\bm 0$. Hence we partition the list according to the distance $\omega = W_H(\bm u_i - \bm u)$. Similar to the proof of Theorem~\ref{main_theorem}, we have
	\begin{equation}
		\begin{aligned}
			{\rm{Pr}}\{{\rm error}|\bm u\}
			&\leq\exp\Big\{-nE\Big(\frac{1}{2},R_M\Big)\Big\}+\sum\limits_{\bm y \in \mathcal{Y}^{m}}  P(\bm y|\bm u)\cdot {\rm {Pr}}\Big\{{\bigcup\limits_{i=1}^LA_{i}\Big\}}\\
			&\leq \exp\Big[-kE\Big(\frac{1}{2},R_M\Big)\Big]+\sum\limits_{\omega=1}^{T-1}\exp\Big[-mE(\rho
			_{\omega},R_\omega)\Big]+\exp\Big[{-mE\Big(\frac{1}{2},{R}_T\Big)}\Big]\text{,}
		\end{aligned}	
	\end{equation}
	where  $R_\omega=\log{\binom{k}{\omega}}/m$ and ${R}_T= \log[1+(1-2\rho)^T]+ {k}(1-C+\delta/2)/m$.  From Theorem~\ref{partial_error_exponent}, we  see that the first term can be made not greater than $\epsilon/3$  for sufficiently large $k$. Now letting $k \rightarrow \infty$ and $T \rightarrow \infty$, we have ${R}_T < 1-{\delta}^\prime $ for some ${\delta}^\prime>0$ since $\rho \leq 1/2$. We have from Theorem~\ref{partial_error_exponent} that $E(1/2, {R}_T) > 0 $ since the channel capacity of the noiseless channel is $H(1/2)=1$ and the third term in the right hand side~(RHS) of the inequality~\eqref{result} can be made not greater than $\epsilon/3$ for sufficiently large $k$ since $m$ increases linearly with $k$. By fixing $T$ and for $\omega < T$,  $\log{\binom{k}{\omega}}$ increases only logarithmically with $k$, we have $R_\omega \rightarrow 0 $ as $k \rightarrow \infty$. Since the channel for transmitting $\bm x^m$ is noiseless, we have $I(\rho_\omega)=H(\rho_\omega)=-(1-\rho_\omega)\log(1-\rho_\omega)-\rho_\omega\log\rho_\omega>0$ and hence $E(\rho_{\omega},R_\omega) > 0$ for sufficiently large $k$, implying that the second term in the RHS of~\eqref{result} can also be made not greater than~$\epsilon/3$.
	Now we have
	\begin{equation}
		{\rm Pr}\{{\rm error}|\bm u\} \leq \epsilon.
	\end{equation}
Therefore, ${\rm Pr}\{{\rm error} \} = \sum_{\bm u^{k} \in \mathbb{F}_2^{k} } 2^{-k} {\rm Pr}\{{\rm error}|\bm u^{k}\} \leq  \epsilon$.
	\par { From the weak law of large numbers, we have, for sufficiently large $k$
	\begin{equation}
		{\rm Pr}\{|S(\mathbf{G})-\rho|\leq\eta\}\geq 1-\eta.
	\end{equation}
	Then we assert that there must exist a matrix $\mathbf{G}$ with $S(\mathbf{G}) \leq \rho+ \eta$ and ${\rm Pr}\{{\rm error}|\mathbf{G}\} \leq \epsilon/(1-\eta)$.  Otherwise, we will have ${\rm Pr}\{{\rm error}\} >(1-\eta) \epsilon/(1-\eta) =\epsilon$.
}This completes the proof of Theorem~\ref{LDPC_theorem}.

\par Theorem~\ref{LDPC_theorem} indeed assures that BPC codes are capacity-achieving. More formally, we have 

\begin{corollary}
	A BPC code ensemble can achieve the capacity of a BIOS channel. 
\end{corollary}
\begin{Proof}
	We only need find ways to tell the decoder the noiseless parity-check vector $\bm y^m = \bm x^m$. This can be readily resolved by imposing a constraint that only those $\bm u$ with $\bm u\mathbf{G}=\bm 0$ can be legally transmitted. The code rate is then given by $R = (k-{\rm rank}(\mathbf{G}))/k \geq  (k-m) / k$. 
\end{Proof}
\textbf{Remark:} From the proof, we see that the BPC codes satisfying $k H(U|V) < mI(X;Y)$ can achieve the capacity~(as $k$ and $m$ increase), where $H(U|V)$ is irrelevant to the codes but $I(X;Y)$ increases as the density $\rho$ increases. As an example, consider the $(d_c, d_r)$-regular LDPC code ensemble over the binary symmetric channels~(BSCs) with crossover probability $p$, where $d_c$ and $d_r$ are the column weight and the row weight of the parity check matrix $\mathbf{H} = \mathbf{G}^T$, respectively. With this setting, we see the condition is reduced to 	
\begin{equation}
	d_r H(p)<d_cH(\rho_{d_r})
\end{equation}
where $\rho_{d_r}=(1-(1-2p)^{d_r})/2$ and $H(\cdot)$ is an entropy function. On one hand, given $d_c$ and~$p$, we can find a lower bound on the row weight $d_r$. On the other hand, given $d_c$ and $d_r$, we can find the upper bound on the  crossover probability $p$. From this, we see that the threshold for $(3, 6)$-regular LDPC code can be upper bounded by $p \approx 0.102$, which is consistent with the result in~\cite[Figure 3.5]{Gallager1963LDPC}.

\section{Finite Length Performance of   Systematic BGM Codes}
\label{sec5}

\subsection{Performance Upper Bounds}

In the proof of Theorem~\ref{main_theorem}, we have upper bounded the ensemble average FER by using error exponents. We can also upper bound the FER and the BER~\cite{2013Ma,Cai2020SCLDGM} by deriving input-output weight enumerating function~(IOWEF)
\begin{equation}
	A(X,Y)=\sum\limits_{i=0}^{k}\sum\limits_{j=0}^{m} A_{i,j}X^i Y^j\text{,}
\end{equation}
 as defined in~\cite{Benedetto1996IRWEF}, where $A_{i,j}$ represents the number of codewords having input~(message bits) weight $i$ and redundancy~(parity-check bits) weight~$j$. Distinguished from most existing capacity-achieving code ensembles, the IOWEF of the systematic BGM code ensemble can be readily given as~\cite{Cai2020SCLDGM}
\begin{equation}
	\label{IOWEF}
	A(X,Y)=\sum\limits_{\omega=0}^{k} \binom{k}{\omega} X^\omega (1-\rho_\omega+\rho_\omega Y)^{m}\text{,}
\end{equation}
where 
\begin{equation}
	\rho_{\omega}=\frac{1-(1-2\rho)^{\omega}}{2}\text{.}
\end{equation}
Notice that even truncated IOWEF can be used to derive improved union bounds~\cite{2013Ma}.
\par Similarly, the weight distribution of the BPC code ensemble  can be easily written as\footnote{This can be obtained by simply setting $Y = 0$ in~\eqref{IOWEF}.} 
\begin{equation}
	A(X)=\sum\limits_{i=0}^kA_iX^i=\sum\limits_{\omega=0}^k\binom{k}{\omega}(1-\rho_\omega)^mX^\omega\text{,}
\end{equation}
where $A_i$ is the average number of codewords with Hamming weight $i$.
\subsection{Proof of Theorem~\ref{theorem_BER_bound}}
  
\textit{Proof of Theorem~\ref{theorem_BER_bound}}:
Let $\omega_i$ be the Hamming weight of the $i$-th row of $[\mathbf{I}~\mathbf{G}]$. The error probability of the $i$-th bit $U_i$ is lower bounded by that with the genie-aided decoder~\cite{Ma2015BMST}, which assumes that all but $U_i$ is known. Therefore, the lower bound is equal to the performance of repetition code with length $\omega_i$.
	Without loss of generality, we assume that the receiving sequence corresponding to the repetition code is $\bm y=(y_0,\cdots,y_{\omega_i-1})$. Upon receiving $\bm y$, the optimal decision with log-likelihood ratio~(LLR)  for the repetition code is given by
	\begin{equation}
		\hat{U_i}=\left\{\begin{aligned}
			&0,\qquad \sum\limits_{j=0}^{\omega_i-1}\frac{2y_j}{\sigma^2}>0\\
			&1,\qquad {\text{\rm otherwise}}
		\end{aligned}\right.\text{.}
		\label{flipping}
	\end{equation}
	For the optimal decision,  the error probability for both bit 0 and bit 1 is $Q(\sqrt{\omega_i}/{\sigma})$.  Hence,  the error probability for the repetition code with length $\omega_i$ is
	\begin{equation}
		\begin{aligned}
			P_{W}(\omega_i)=& Q\Big(\frac{\sqrt{\omega_i}}{\sigma}\Big).
		\end{aligned}
	\end{equation}
By averaging the lower bound on BER of $k$ bits, we complete the proof.



\subsection{Proof of Theorem~\ref{theorem_FER_bound}}
\textit{Proof of Theorem~\ref{theorem_FER_bound}:}
	Denote by $B_i$ the event that $\{P(\bm y|\bm s_0)\geq P(\bm y|\bm s_i)\}$. The event $B_i$ occurs when $\bm y$ is closer to $\bm s_0$ than $\bm s_i$ in $\mathbb{R}^{k+m}$. Thus, given that the Hamming weight of $\bm c_i$ is $\omega_i$, we have

	\begin{equation}
		{\rm Pr}\{B_i\}\leq 1-Q\Big(\frac{\sqrt{\omega_i}}{\sigma}\Big)\text{.}
	\end{equation}
	 Now setting  $\tilde{\Omega}=\{\cap_{i=1}^L {B_i}\}$, we have  
	\begin{equation}
		\begin{aligned}
		{\rm Pr}\{\bm y\in \Omega\}&\leq{\rm Pr}\{\bm y\in \tilde{\Omega}\}\\
			&=	{\rm Pr}\Big\{\bigcap\limits_{i=1}^L {B_i}\Big\}\\
			&\overset{(*)}{=}\prod\limits_{i=1}^L{\rm Pr}\{{B_i}\} \\
			&\leq \prod\limits_{i=1}^L\Big[1-Q\Big(\frac{\sqrt{\omega_i}}{\sigma}\Big)\Big]\text{.}
		\end{aligned}
	\end{equation}
where $(*)$ follows from the assumption that  $\overrightarrow{\bm s_0\bm s_i}~(1\leq i\leq L)$ are orthogonal to  each other.
	Hence, we have 
	\begin{equation}
		{\rm FER}=1-{\rm Pr}\{\bm y\in \Omega\}\geq 1-\prod_{i=1}^{L}\Big[1-Q\Big(\frac{\sqrt{\omega_i}}{\sigma}\Big)\Big]\text{,}
	\end{equation}
and completes the proof.
\par \textbf{Remarks:} Notice that Theorems~\ref{theorem_BER_bound} and~\ref{theorem_FER_bound} also apply to general linear block codes over BPSK-AWGN channels. Also notice that, for  systematic BGM codes, we may consider the following approximate lower bound
	\begin{equation}
	{\rm FER}\gtrsim 1-\prod_{i=1}^{k}\Big[1-Q\Big(\frac{\sqrt{\omega_i}}{\sigma}\Big)\Big]\text{,}
\end{equation}
where $\omega_i$ is the Hamming weight of the $i$-th row of $[\mathbf{I}~\mathbf{G}]$. This  approximately holds since the rows of  $[\mathbf{I}~\mathbf{G}]$ with sparse $\mathbf{G}$ are {non-overlap~(orthogonal with BPSK signalling) or near non-overlap to each other with high probability.}

\subsection{Numerical Results}

\par
\begin{figure}[tbp]     
	\centering     
	\subfigure[The BER performance.]{           
		\centering  
		\label{BER}   
		\includegraphics[width=0.6\textwidth]{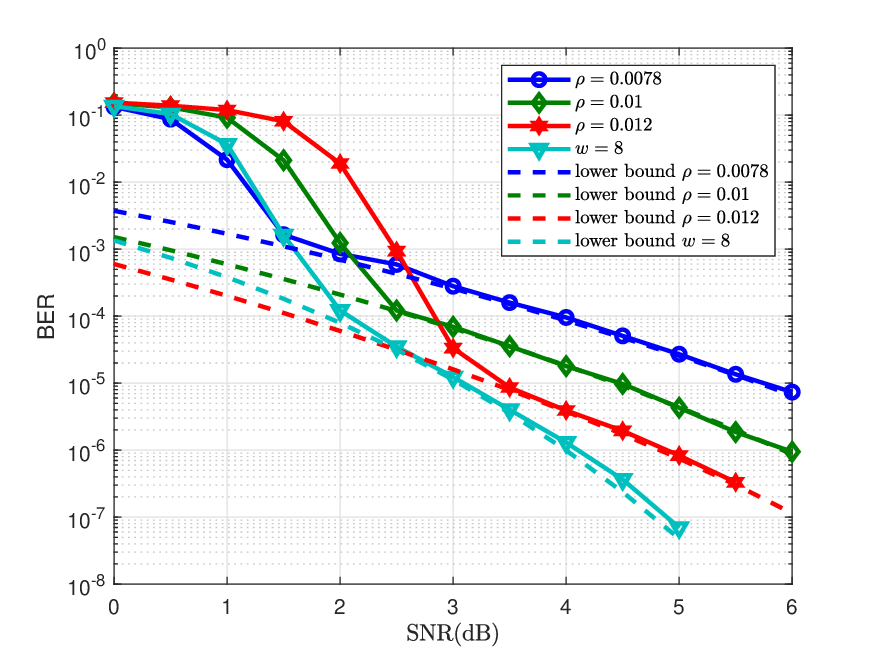}         }    
	\subfigure[The FER performance.]{         
		\centering   
		\label{FER}   
		\includegraphics[width=0.6\textwidth]{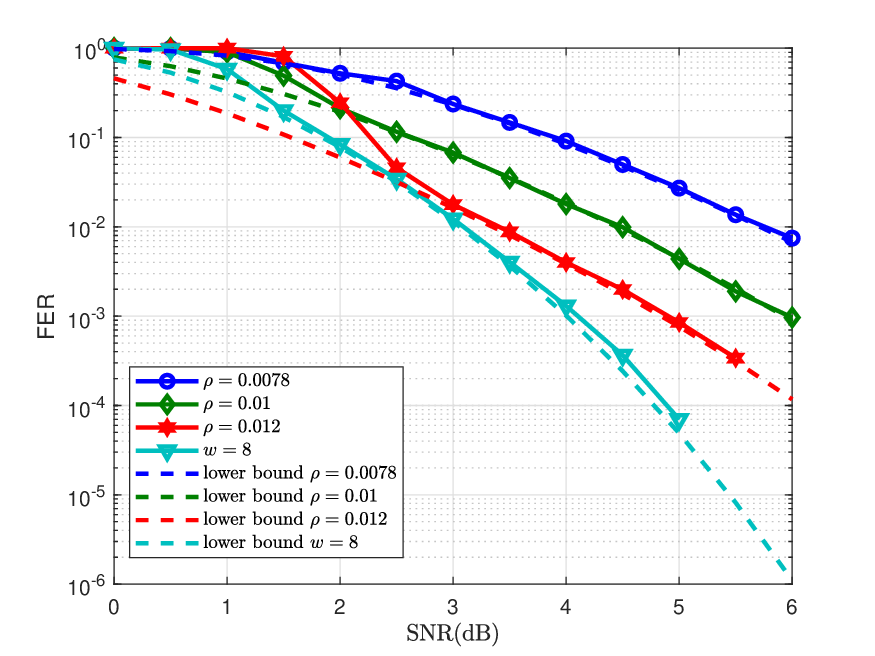}          }   
	\caption{The BER and FER performance of  systematic BGM codes with $k=1024$ and $m=1024$. The parameter $\rho$ is the density of the Bernoulli random matrix $\mathbf{G}$ and ``$w = 8$'' is for the fixed row weight of a random matrix $\mathbf{G}$ constructed by Gallager's approach. } 
\end{figure}

 \begin{example}
	Consider  systematic BGM code with $k=1024$, $m=1024$.   The density  $\rho$ of $\mathbf{G}$ is specified in the legends. The simulation results~(with iterative belief propagation~(BP) decoding algorithm instead of the infeasible ML decoding algorithm) and the derived lower bounds are shown in Fig.~\ref{BER} and Fig.~\ref{FER}. We  see from the figures that, in the high SNR region, the BER and FER performance of  the systematic BGM codes can match well with the corresponding lower bounds. As predicted by the lower bounds, which are decreasing with the row weights of $\mathbf{G}$, the higher density of $\mathbf{G}$ is, the lower error floors are. For $\rho=0.0078$, the average row weight of $\mathbf{G}$ is about~$8$.  However, randomly generating the matrix cannot guarantee the minimum row weight of $\mathbf{G}$. As a result, the performance with a Bernoulli matrix $\mathbf{G}$ of density $\rho = 0.0078$ is worse than that with a random matrix of fixed row weight $\omega = 8$. Notice that the random construction of $\mathbf{G}$ with fixed row weight is similar to Gallager's construction of the low density parity check matrices~\cite{1962Gallager}. 
	
\end{example} 

\section{A Statistical Physics Approach to Optimizing BGM Codes}
\label{sec6}
\par It is commonly accepted that the performance curves of iteratively decodable codes can be  roughly divided into the water-fall region and the error-floor region. The error-floor region can be improved by concatenating with tailored outer codes, while the water-fall region can be improved by optimizing the degree distributions with the extrinsic information transfer (EXIT) chart analysis~\cite{2001EXIT}. Distinguished from most existing works, we investigate the impact of node connection preferences on the performance for the BGM codes, whose degree distributions are fixed as binomial distributions. To this end, we turn to the statistical physics approaches following the work of Sourlas~\cite{sourlas1989spin}. We first introduce the quantitative metrics for characterizing spin interaction patterns. Then, we propose the bipartite graph configuration model to generate
BGM codes with targeted assortativity coefficients. Finally, we compare the iterative BP decoding
performance of those BGM codes, and predict the asymptotic performance of the disassortative
BGM codes ensemble by population dynamics.

\subsection{Assortativity Coefficient of Bipartite Graph}
 In the complex network theory, the assortativity coefficient is employed to quantify the tendency of nodes within a network to be connected to other nodes with similar or different degrees~\cite{Newman2002PRL}. Here, we specialize the assortativity coefficients to the bipartite graphs associated with the BGM codes.

Given a bipartite graph with $N$~(in total) nodes and $M$ edges, we denote by $N_{j}$ the number of nodes with degree $j$, and by $M_{ij}$ the number of edges that connect nodes with degree $i$ to nodes with degree $j$. Then we define the degree distribution as
\begin{equation}
	p_{j} = \frac{N_{j}}{N},
\end{equation}
and the joint degree distribution as
\begin{equation}
	e_{ij} = \frac{M_{ij}}{M}.
\end{equation}
That is to say, if a node is selected uniformly  at random, the probability of its degree being $j$ is $p_{j}$; if an edge is selected uniformly at random, the probability of its end nodes having degrees $i$ and $j$ is $e_{ij}$.
Next, we define the excess degree distribution as 
\begin{equation}
	q_{j} = \frac{j p_{j}}{\sum_{j^\prime} j^\prime p_{j^\prime}},
\end{equation}
where $\sum_{j^\prime} j^\prime p_{j^\prime}$ is the average degree. It means that the probability of reaching a node with degree-$j$ along a randomly selected edge is $q_j$. Similar to the original definition~\cite{Newman2002PRL}, we now define the assortativity coefficient of bipartite graphs as 
\begin{equation}\label{eq:AssCoef}
	r = \frac{1}{\sigma_{q}^{2}} \sum_{ij} ij(e_{ij}-q_{i}q_{j}),
\end{equation}
where $\sigma_{q}^{2}$ is variance of the distribution $q_{j}$, given by
\begin{equation*}
	\sigma_{q}^{2} = \sum_{j} j^{2} q_{j} - \left[ \sum_{j} j q_{j} \right]^{2}.
\end{equation*}
The assortativity coefficient $r$ ranges from $-1$ to $+1$. 
Real networks typically exhibit different assortativity coefficients and can be classified into three types of networks, namely, disassortative~($r<0$), neutral($r \approx 0$), and assortative~($r>0$)~\cite{Barabasi2016NS}. 
For instance, many metabolic networks are disassortative, where molecules with high degrees often connect to those with low degrees~(see Fig.~\ref{fig:BGM_Eija} top). 
In contrast, coauthorship networks are often assortative, suggesting that authors tend to collaborate with individuals who have a similar number of connections~(see Fig.~\ref{fig:BGM_Eijc} top). 
Some other networks, such as the power grid, are neutral, implying that the connections between nodes are random~(see Fig.~\ref{fig:BGM_Eijb} top).
\subsection{Bipartite Graph Configuration Model}
Based on the configuration model in network science~\cite{Barabasi2016NS}, we propose the bipartite graph configuration model with specific assortativity coefficient.
Define the probability of a variable node with degree  $k_{v}$ connecting to a check node with degree $k_{c}$ as 
\begin{equation}\label{eq:JointDegDist}
	P(k_{v}, k_{c}, a) = \frac{1}{Z} |(k_{v}-\bar{k}_{v}) - (k_{c}-\bar{k}_{c})| ^ a, 
\end{equation}
where $\bar{k}_{v}$ is the average degree of the variable nodes and $\bar{k}_{c}$ is the average degree of check nodes, $Z$ is the normalized factor, and $a$ is the parameter that controls the assortativity coefficient. 
This model takes as input variable nodes' degree sequence $D_{1}$, check nodes' degree sequence $D_{2}$, targeted $r^{*}$, tolerance error $\epsilon$, $a_{1}$ and $a_{2}$, and delivers as output the adjacent matrix with targeted $r^{*}$ by performing a binary search on the parameter $a$. Let $S_{1}$ and $S_{2}$ be the variable-stub and check-stub lists. 
Let $T_{\max}$ be the maximum number of attempts allowed for randomly selecting two stubs and successfully establishing an edge between them. 
The concrete procedure is summarized in Algorithm~\ref{alg:NGCM}. 
Notice that for systematic BGM codes with $k=m$, Algorithm~\ref{alg:NGCM} can generate normal graphs with either $r^{*}>0$ or $r^{*}<0$. 
However, for systematic BGM codes with $k \neq m$, only normal graphs with $r^{*}<0$ can be generated due to their inherent disassortativity. 

\begin{figure}[t]
	\centering
	\subfigure[disassortative]{           
		\centering  
		\includegraphics[width=0.31\textwidth]{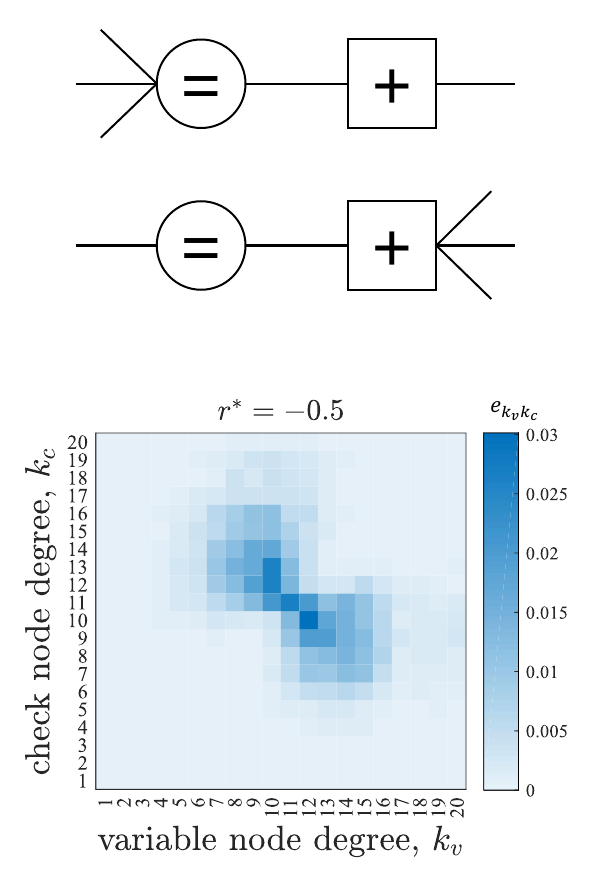}   
		\label{fig:BGM_Eija}      }    
	\subfigure[neutral]{         
		\centering    
		\includegraphics[width=0.31\textwidth]{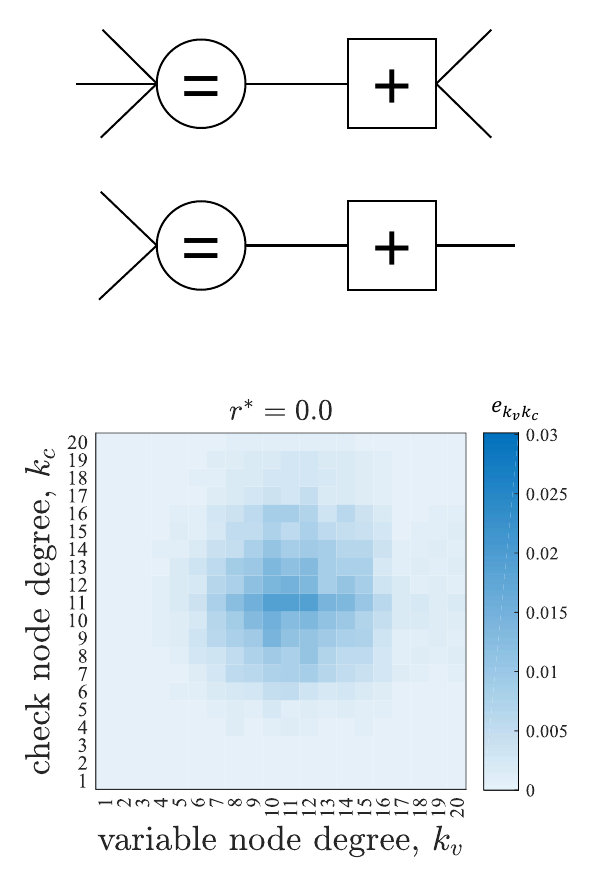}       
			\label{fig:BGM_Eijb}   }   
	\subfigure[assortative]{         
	\centering   
	\label{fig:BGM_Eijc}  
	\includegraphics[width=0.31\textwidth]{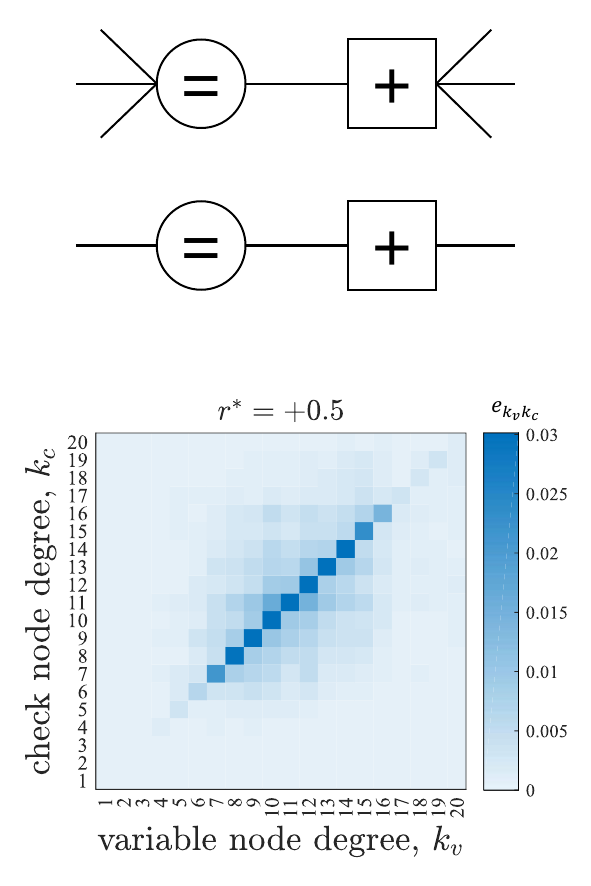}       
   } 
	\caption{An illustration of different node interaction patterns. Here, we take the BGM code with $k=1024, m=1024, \rho=0.01$ as an example. Top: the disassortative~(a), random~(b), and assortative~(c) node interaction patterns. Bottom: the corresponding joint degree distributions.}
	\label{fig:BGM_Eij}
\end{figure}

Fig.~\ref{fig:BGM_Eij} shows the joint degree distributions of three different systematic BGM codes whose normal graph is generated by Algorithm~\ref{alg:NGCM} for a given $r^{*}$. 
We can observe that, for $r^*=-0.5$, nodes with larger degree differences have a higher probability of connection, while for $r^{*}=+0.5$, nodes with similar degrees have a higher probability of connection. 
For $r^*=0.0$, the probability of connection is homogeneous. 


\subsection{Performance of the Disassortative Systematic BGM Codes}
Here, we first compare and analyze the iterative BP decoding performance of systematic BGM codes with different assortativity coefficients. Then, we explain intuitively the coding gains from a complex network perspective. Finally, we present the asymptotic performance predicted by the population dynamics.

\begin{figure}[t]
	\centering
	\includegraphics[width=0.65\textwidth]{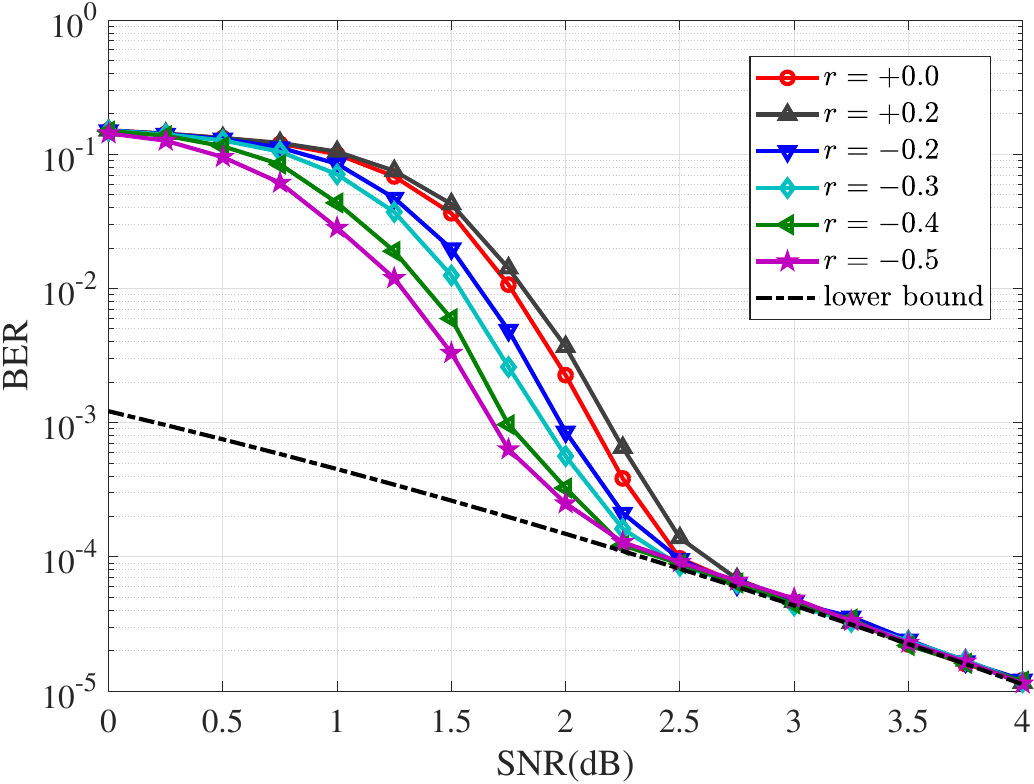}
	\caption{The BER performance of systematic BGM codes of $k=1024$, $m=1024$ and $\rho=0.01$ with different assortativity coefficient $r$. The black dashed line represents the lower bound. }
	\label{fig:BGM_BER}
\end{figure}

\begin{figure}[t]
	\centering
	\includegraphics[width=0.65\textwidth]{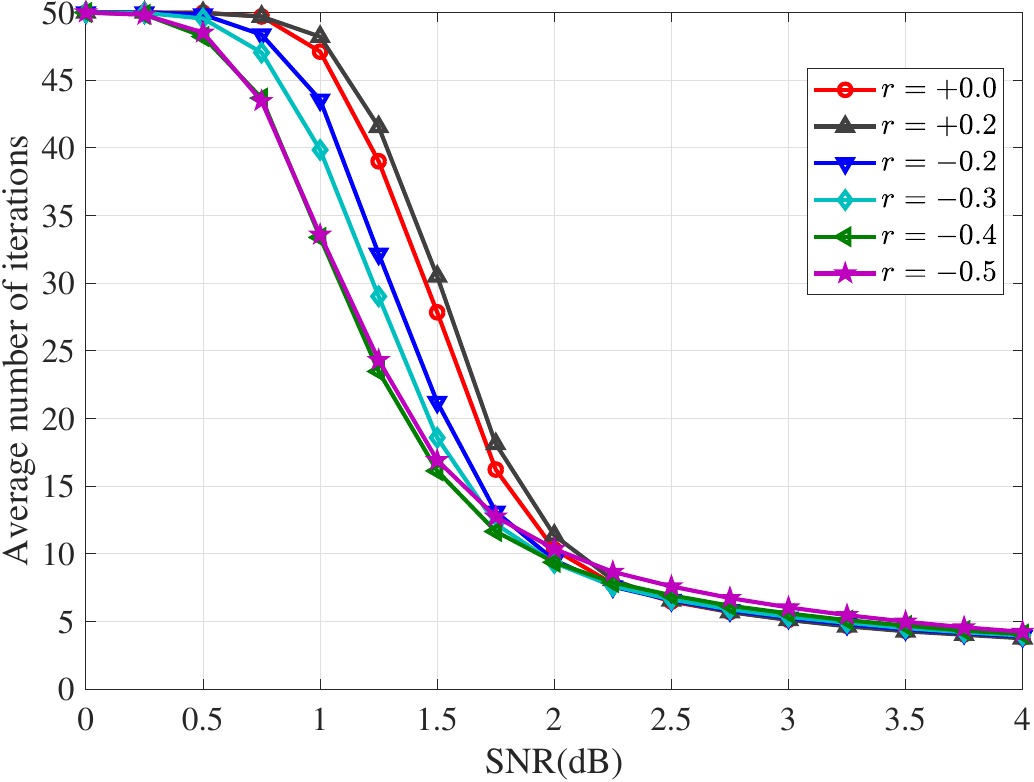}
	\caption{Average number of iterations of systematic BGM codes of $k=1024$, $m=1024$ and $\rho=0.01$ with different assortativity coefficient $r$. }
	\label{fig:BGM_Iter}
\end{figure}

\begin{example}
	 Consider systematic BGM code with $k=1024, m=1024, \rho=0.01$. Assume that the codeword is mapped to BPSK signals and transmitted over AWGN channels. Firstly, we utilize Algorithm~\ref{alg:NGCM} to generate disassortative normal graphs with different assortativity coefficient~$r$. Then, we perform the sum-product algorithm for decoding, where the maximum iteration number is set to be $50$. The simulation results are shown in Fig.~\ref{fig:BGM_BER} and Fig.~\ref{fig:BGM_Iter}. We can observe that the assortative BGM code~(with $r=+0.2$) exhibits degraded BP performance compared to the original code~($r=+0.0$), whereas disassortativity~($r<0$) leads to improved performance in the waterfall region. Specifically, as $r$ decreases further, the coding gains are increasing, with about $0.5$ dB for $r=-0.50$ at BER of $10^{-3}$. In addition, the decoding complexity decreases for $r=-0.50$ in the water-fall region as well.
\end{example}
\begin{algorithm}[!htb]
	\setstretch{1}
	\caption{Bipartite Graph Configuration Model with Specific Assortativity Coefficient}\label{alg:NGCM}
	\small
	\KwIn{$D_{1}$, $D_{2}$, $T_{\max}$, $r^{*}$,  $\epsilon$, $a_{1}$, $a_{2}$}
	\textbf{Initialization}: $\tilde{r}=0$ \\
	\While{$|\tilde{r} - r^{*}| > \epsilon$} {
		${\rm flag}=1$\;
		$a = (a_{1}+a_{2})/2$\;
		generate $P(k_{v}, k_{c}, a)$ according to~\eqref{eq:JointDegDist}; \\
		\If{$r^{*} > 0$} {
			$P(k_{v}, k_{c}, a) = 1 -P(k_{v}, k_{c}, a)$\;
		}
		\While{${\rm flag}==1$} {
			$\mathbf{A}=\mathbf{0}$, ${\rm flag}=0$\;
			$S_{1}=[~]$, $S_{2}=[~]$\;
			add $D_{1}(v)$'s stubs $v$ into $S_{1}$ for each $v$. \\
			add $D_{2}(c)$'s stubs $c$ into $S_{2}$ for each $c$. \\
			\While{$|S_{1}| > 0$ {\rm and} $|S_{2}| > 0$} {
				$t = 0$\;
				\While{$t < T_{\max}$} {
					randomly select a stub $v$ from $S_{1}$\;
					randomly select a stub $c$ from $S_{2}$\;
					\If{$\mathbf{A}(c, v) == 0$} {
						$k_{v} = D_{1}(v)$, $k_{c} = D_{2}(c)$\;
						generate random number $p \in (0, 1)$\;
						\If{$p \leq P(k_{v}, k_{c}, a)$} {
							$\mathbf{A}(c, v) = 1$\;
							delete $v$ from $S_{1}$\;
							delete $c$ from $S_{2}$\;
							\textbf{break}\;
						}
					}
					$t = t + 1$\;
				}
				\If{$t \geq T_{\max}$} {
					${\rm flag} = 1$\;
					break\;
				}
			}
		}
		calculate $\tilde{r}$ by~\eqref{eq:AssCoef}; \\
		\eIf{$\tilde{r} > r^{*}$} {
			$a_{1} = a$\;
		}{
			$a_{2} = a$\;
		}
	}
	\KwOut{The adjacent matrix $\mathbf{A}$ with targeted $r^{*}$. } 
	
\end{algorithm}

\par
\begin{figure}[t]
	\centering
	\includegraphics[width=0.8\textwidth]{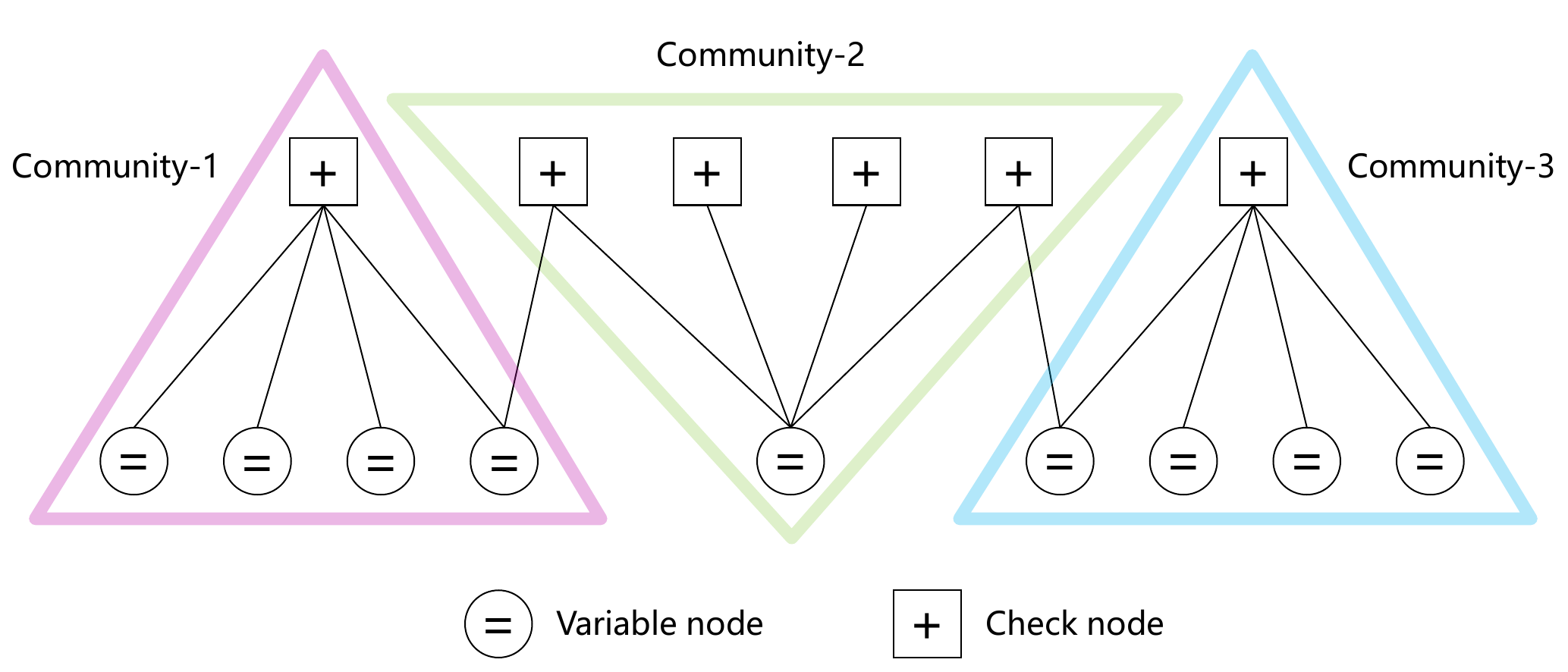}
	\caption{An illustration of community structures in the disassortative normal graph. There are three communities: two with low reliability~(community-1 and -3), one with high reliability~(community-2). }
	\label{fig:NG_Cluster}
\end{figure}

 \textbf{Remarks:} The disassortative gain can be attributed to the network topology, which plays an important role in information spreading. For instance, small-networks, which possess both short average path lengths and high clustering coefficient, have advantages in terms of information dissemination efficiency and speed~\cite{Watts1998NAT}. Another significant finding is that scale-free networks~\cite{Barabasi1999SCI}, whose degree distribution following power-law, can lower down the spreading threshold and expand the spreading scale~\cite{Pastor2001PRL}. Under the scenario of iterative decoding on the disassortative normal graph, there are two essential observations: i) the larger~(smaller) the degree of a variable~(check) node, the higher~(lower) its reliability; ii) disassortativity tends to connect nodes with larger degree differences, thereby forming communities that are more sparse internally and exhibit varying reliability, as shown in Fig.~\ref{fig:NG_Cluster}. From the perspective of transferring extrinsic information, the coding gain in the waterfall region stems from the following two aspects: i) the sparsity within local communities enhances the efficiency of extrinsic information transferring and updating; ii) the high reliability communities~(community-2) further enhance decoding performance, while the edges between these communities facilitate the decoding of low reliability communities~(community-1 and -3) by transferring high-reliability extrinsic information. However, it should be noted that excessive disassortativity hinders the transmission of extrinsic information between communities, thus potentially degrading decoding performance.

To predict the asymptotic performance of disassortative systematic BGM codes, we turn to the population dynamics (also known as sampled density evolution in coding theory)~\cite{Mezard2009information}. Here, we slightly modify the population dynamics by predetermining the degree values at both ends of each edge~(i.e., individual of the population) according to~\eqref{eq:JointDegDist} to adapt the degree-correlated case. 

\begin{figure}[t]
	\centering
	\includegraphics[width=0.65\textwidth]{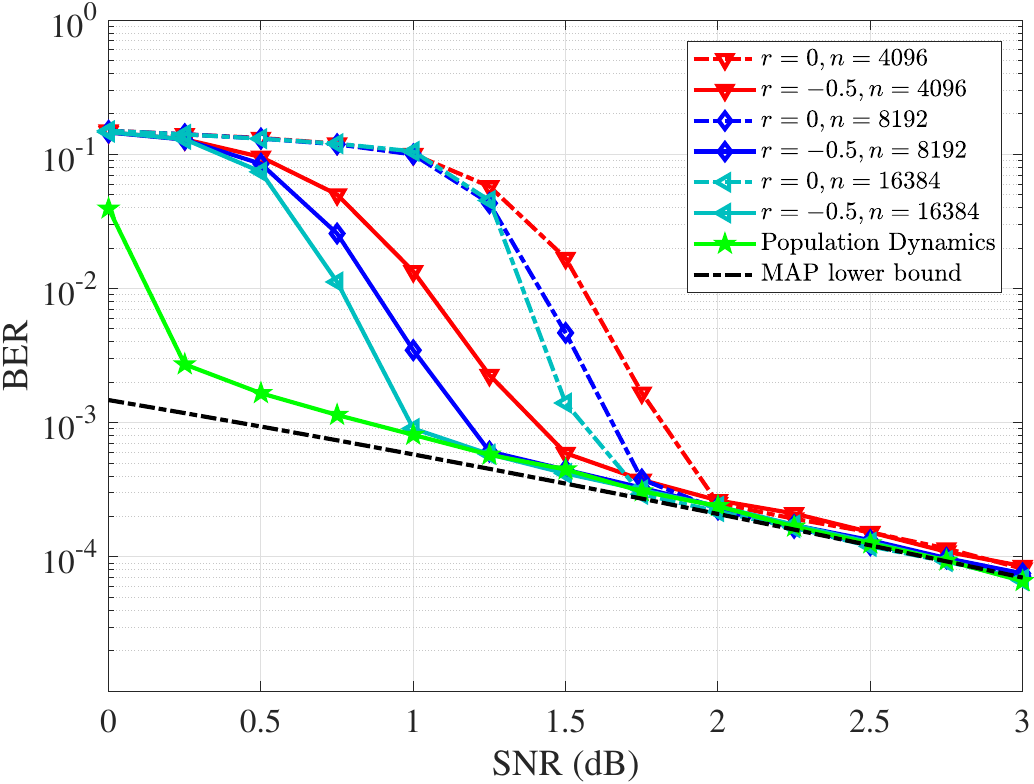}
	\caption{The asymptotic performance of the disassortative systematic BGM codes ensemble. The dashed-line  denote the original~($r=0$) and solid-line markers denote disassortative~($r=-0.5$) systematic BGM codes with different $n$. The green pentagram represents the performance of population dynamics with population size $10^{5}$. The black dashed line represents the MAP lower bounds. }
	\label{fig:Asymptotic}
\end{figure}

\begin{example}
Consider the disassortative systematic BGM codes with code rate $R=1/2$, the optimal assortativity coefficient $r^{*}=-0.5$, and the control parameter $a=2.6$. 
By varying the code length $n=k+m$ and Bernoulli parameter $\rho$, we keep the average degree of variable nodes $\bar{k}_{v}=10.24$. 
Fig.~\ref{fig:Asymptotic} demonstrates the BP performance of the original and disassortative systematic BGM codes with different $n$. 
We can observe that the BP performance in the waterfall region can be significantly improved by utilizing the disassortative systematic BGM codes. 
As the code length $n$ increases, the BP performance gradually approaches the population dynamics's performance. 
In addition, the performance of population dynamics can approaches the MAP lower bound at the low SNR~(about $0.5$ dB). 
\end{example}

\begin{figure}[t]
	\centering
	\includegraphics[width=0.65\textwidth]{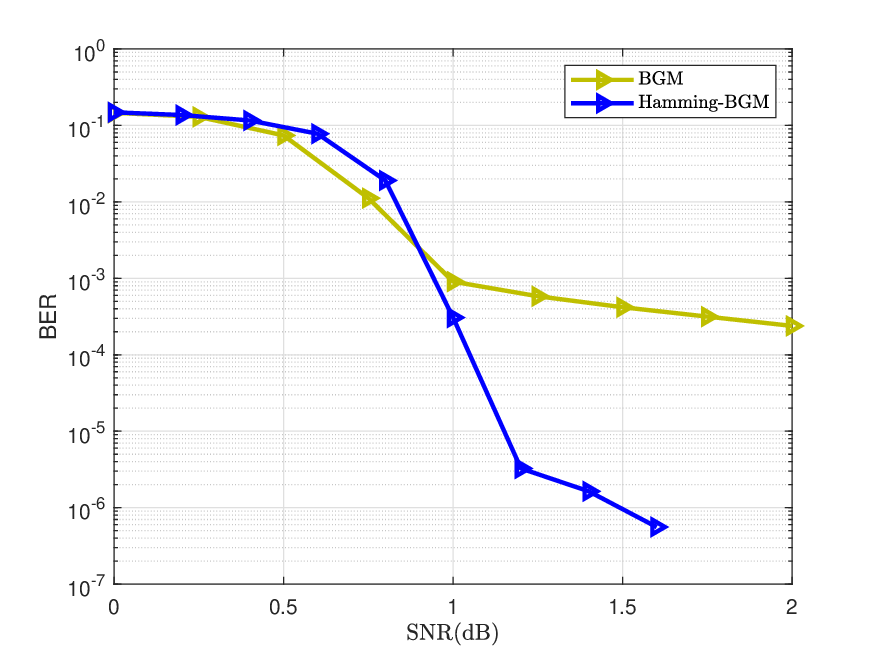}
	\caption{The BER performance of the disassortative systematic BGM code with $k=8192$, $m=8192$ and the outer extended Hamming code concatenated inner code disassortative systematic BGM code with the total code rate $1/2$. }
	\label{Hamming_LDGM}
\end{figure}

\begin{example}
	
We consider a concatenated code consisting of $8$ outer extended Hamming codes $\mathscr{C}[1024,1003]$ and an inner disassortative systematic BGM code with $k=8192$, $m=7856$ and $r^{*}=-0.5$. Notice that  the total rate is  still $1/2$. For the decoding, the BP algorithm is utilized for inner systematic BGM codes and the BCJR algorithm~\cite{1974BCJR} is utilized for the outer Hamming codes.  The BER performance is shown in Fig.~\ref{Hamming_LDGM}, where the performance of the disassortative systematic BGM code with $k=8192$ and $m=8192$ is also plotted.  From the figure, we see that, compared with the disassortative systematic BGM code with $k=8192$ and $m=8192$, concatenating with outer extended Hamming codes can significantly improve the error-floor region performance.  
\end{example}

\section{Conclusions}
\label{sec7}
In this paper, we have proved that both BGM codes and BPC codes  are capacity-achieving over BIOS channels in terms of FER. These two classes of codes with proper designs can be reduced to special classes of LDGM codes and LDPC codes, respectively.  The  proposed framework for proof distinguishes the error exponents for message vectors of different weights and is powerful, which can also be applicable to proving that  the time-invariant random convolutional codes are capacity-achieving in terms of the first error event probability. More generally, this method can be used to prove that  the BER of some bits~(if not all) in linear block codes can be arbitrarily small~\cite{2022FR}. For systematic BGM codes in the finite length region, we derive the lower bounds on the BER and FER to predict the error floors. Numerical results show that the  systematic BGM codes  match well with the derived error floors. The systematic BGM codes suffer from poor water-fall region performance and high error floors. We use the statistical physics approaches to improve the water-fall region performance and then use the Hamming code as outer code to improve the error-floor region performance.

\section*{Appendix}
\subsection{Proof of Lemma~\ref{lemma_for_rho}}
Since the elements of $\mathbf{G}$ are sampled from a Bernoulli process with success probability $\rho$, the parity-check vector $\bm x=\bm u\mathbf{G}$ must be a segment of a Bernoulli process. We only need determine the success probability $\rho_\omega$. This can be solved by induction with $\omega$ as shown in~\cite[Lemma 1]{Cai2020SCLDGM} or  by analysis provided in~\cite[Lemma 1]{1962Gallager}. 

\par Without loss of generality, we may assume that the information vector $\bm u$ with weight $\omega$ has the form of $(\bm 1^\omega ~\bm 0^{k-w})$, consisting of $\omega$ ones followed by $k-\omega$ zeros. The  $j$-th  parity-check bit is $X_j = \sum_{i=0}^{\omega} G_{ij}$, a module-$2$ sum of $\omega$ Bernoulli variables. The event $\{X_j=1\}$ is equivalent to that an odd number of ones appear in $\{G_{i,j}\}$, whose probability is given by 
\begin{equation}
	{\rm Pr}\{X_j=1|W_H(\bm u)=\omega\}=\sum\limits_{0\leq \ell\leq \omega,~\ell \text{~is odd}  }\binom{\omega}{\ell }\rho^{\ell}(1-\rho)^{\omega-\ell}\text{.}
\end{equation}
By expanding $(1-2\rho)^\omega$ as $\sum_{\ell=0}^{\omega} \binom{\omega}{\ell}(1-\rho)^{\omega-\ell} (-\rho)^\ell$, we see that
\begin{equation}
	(1-2\rho)^\omega = {\rm Pr}\{X_j = 0|W_H(\bm u)=\omega\} -  {\rm Pr}\{X_j = 1|W_H(\bm u)=\omega\}\text{.}
\end{equation}
Therefore, from ${\rm Pr}\{X_j = 0|W_H(\bm u)=\omega\}+{\rm Pr}\{X_j = 1|W_H(\bm u)=\omega\}=1$, we have
\begin{equation}
	\rho_\omega=\frac{1-(1-2\rho)^{\omega}}{2}\text{.}
\end{equation}
Noticing that $\rho\leq \rho_\omega\leq \rho_{\omega+1}\leq 1/2$, we have $P(\bm x^{m}|\bm u)\leq P(\bm 0^m|\bm u)\leq (1-\rho_T)^m$ for all $\bm u\in \mathbb{F}_2^k$ with $W_H(\bm u)\geq T$ and $\bm x\in \mathbb{F}_2^m$.
\section*{Acknowledgment}
The authors would like to thank Dr. Suihua Cai from Sun Yat-sen University  
for his helpful discussions.

\bibliographystyle{IEEEtran}
\bibliography{bibliofile}

\begin{thebibliography}{10}
\providecommand{\url}[1]{#1}
\csname url@samestyle\endcsname
\providecommand{\newblock}{\relax}
\providecommand{\bibinfo}[2]{#2}
\providecommand{\BIBentrySTDinterwordspacing}{\spaceskip=0pt\relax}
\providecommand{\BIBentryALTinterwordstretchfactor}{4}
\providecommand{\BIBentryALTinterwordspacing}{\spaceskip=\fontdimen2\font plus
\BIBentryALTinterwordstretchfactor\fontdimen3\font minus
  \fontdimen4\font\relax}
\providecommand{\BIBforeignlanguage}[2]{{%
\expandafter\ifx\csname l@#1\endcsname\relax
\typeout{** WARNING: IEEEtran.bst: No hyphenation pattern has been}%
\typeout{** loaded for the language `#1'. Using the pattern for}%
\typeout{** the default language instead.}%
\else
\language=\csname l@#1\endcsname
\fi
#2}}
\providecommand{\BIBdecl}{\relax}
\BIBdecl

\bibitem{Gallager1968}
R.~Gallager, \emph{Information Theory and Reliable Communication}.\hskip 1em
  plus 0.5em minus 0.4em\relax New York, NY: John Wiley and Sons, Inc., 1968.

\bibitem{Elias1955Coding}
P.~Elias, ``Coding for noisy channels,'' \emph{IRE Conv. Rec.}, vol.~4, pp.
  37--46, 1955.

\bibitem{1962Gallager}
R.~Gallager, ``Low-density parity-check codes,'' \emph{IRE~Trans.~Inf.~Theory},
  vol.~8, no.~1, pp. 21--28, Jan. 1962.

\bibitem{Richardson2001DE}
T.~Richardson and R.~Urbanke, ``The capacity of low-density parity-check codes
  under message-passing decoding,'' \emph{IEEE Trans. Inf. Theory}, vol.~47,
  no.~2, pp. 599--618, Feb. 2001.

\bibitem{Richardson2001LDPC}
T.~Richardson, M.~Shokrollahi, and R.~Urbanke, ``Design of capacity-approaching
  irregular low-density parity-check codes,'' \emph{IEEE Trans. Inf. Theory},
  vol.~47, no.~2, pp. 619--637, Feb. 2001.

\bibitem{MacKay1999sparse}
D.~MacKay, ``Good error-correcting codes based on very sparse matrices,''
  \emph{IEEE Trans. Inf. Theory}, vol.~45, no.~2, pp. 399--431, Mar. 1999.

\bibitem{sourlas1989spin}
N.~Sourlas, ``Spin-glass models as error-correcting codes,'' \emph{Nature},
  vol. 339, no. 6227, pp. 693--695, Jun. 1989.

\bibitem{cheng1996some}
J.-F. Cheng and R.~J. McEliece, ``Some high-rate near capacity codecs for the
  {Gaussian} channel,'' in \emph{Proc. 34th Allerton Conf. Commun., Control
  Comput.}, vol.~34, Oct. 1996, pp. 494--503.

\bibitem{Luby2002LT}
M.~Luby, ``{LT} codes,'' in \emph{Annu. IEEE Symp. Found. Comput. Sci.},
  Vancouver, Canada, Nov. 2002, pp. 271--280.

\bibitem{Montanari2005Tight}
A.~Montanari, ``Tight bounds for {LDPC} and {LDGM} codes under {MAP}
  decoding,'' \emph{IEEE Trans. Inf. Theory}, vol.~51, no.~9, pp. 3221--3246,
  Sept. 2005.

\bibitem{2003Approaching}
J.~Garcia-Frias and W.~Zhong, ``Approaching {Shannon} performance by iterative
  decoding of linear codes with low-density generator matrix,'' \emph{IEEE
  Commun. Lett.}, vol.~7, no.~6, pp. 266--268, Jun. 2003.

\bibitem{Lopez2007Serially}
M.~Gonzalez-Lopez, F.~J. Vazquez-Araujo, L.~Castedo, and J.~Garcia-Frias,
  ``Serially-concatenated low-density generator matrix {(SCLDGM)} codes for
  transmission over {AWGN} and {Rayleigh} fading channels,'' \emph{IEEE Trans.
  Wireless Commun.}, vol.~6, no.~8, pp. 2753--2758, Aug. 2007.

\bibitem{Shokrollahi2006Raptor}
A.~Shokrollahi, ``Raptor codes,'' \emph{IEEE Trans. Inf. Theory}, vol.~52,
  no.~6, pp. 2551--2567, Jun. 2006.

\bibitem{Wainwright2007}
M.~J. Wainwright, ``Sparse graph codes for side information and binning,''
  \emph{IEEE Signal Process. Mag.}, vol.~24, no.~5, pp. 47--57, Sept. 2007.

\bibitem{Zhu2021Compression}
T.~Zhu and X.~Ma, ``Near-lossless compression for sparse source using
  convolutional low density generator matrix codes,'' in \emph{Proc.~Data
  Compression Conf.~(DCC)}, Mar. 2021, pp. 323--332.

\bibitem{Wainwright2010}
M.~J. Wainwright, E.~Maneva, and E.~Martinian, ``Lossy source compression using
  low-density generator matrix codes: Analysis and algorithms,''
  \emph{IEEE~Trans.~Inf.~Theory}, vol.~56, no.~3, pp. 1351--1368, March. 2010.

\bibitem{Golmohammadi2018}
A.~Golmohammadi, D.~G.~M. Mitchell, J.~Kliewer, and D.~J. Costello, ``Encoding
  of spatially coupled {LDGM} codes for lossy source compression,'' \emph{IEEE
  Trans. Commun.}, vol.~66, no.~11, pp. 5691--5703, Nov. 2018.

\bibitem{Alinia2023}
M.~Alinia and D.~G.~M. Mitchell, ``Optimizing parameters in soft-hard {BPGD}
  for lossy source coding,'' in \emph{12th Int. Symp. Top. Coding (ISTC)},
  Sept. 2023, pp. 1--5.

\bibitem{Etesami2006}
O.~Etesami and A.~Shokrollahi, ``Raptor codes on binary memoryless symmetric
  channels,'' \emph{IEEE Trans. Inf. Theory}, vol.~52, no.~5, pp. 2033--2051,
  May 2006.

\bibitem{Shirvanimoghaddam2016}
M.~Shirvanimoghaddam and S.~Johnson, ``Raptor codes in the low {SNR} regime,''
  \emph{IEEE Trans. Commun.}, vol.~64, no.~11, pp. 4449--4460, Nov. 2016.

\bibitem{Kharel2018}
A.~Kharel and L.~Cao, ``Analysis and design of physical layer {Raptor} codes,''
  \emph{IEEE Commun. Lett.}, vol.~22, no.~3, pp. 450--453, Mar. 2018.

\bibitem{Zhang2017}
L.~M. Zhang and F.~R. Kschischang, ``Low-complexity soft-decision concatenated
  {LDGM}-staircase {FEC} for high-bit-rate fiber-optic communication,''
  \emph{J. Lightwave Technol.}, vol.~35, no.~18, pp. 3991--3999, Sept. 2017.

\bibitem{Alinia2024}
M.~Alinia and D.~G.~M. Mitchell, ``Minimizing distortion in data embedding
  using {LDGM} codes and the cavity method,'' in \emph{Proc.~
  Int.~Symp.~Inf.~Theor.~(ISIT)}, Jul. 2024, pp. 226--231.

\bibitem{Yao2024}
Q.~Yao, W.~Zhang, K.~Chen, and N.~Yu, ``{LDGM} codes-based near-optimal coding
  for adaptive steganography,'' \emph{IEEE Trans. Commun.}, vol.~72, no.~4, pp.
  2138--2151, Apr. 2024.

\bibitem{Ma2016Coding}
X.~Ma, ``Coding theorem for systematic low density generator matrix codes,'' in
  \emph{Proc.~9th~Int. Symp. Turbo Codes Iterative Inf. Process.~(ISTC)}, Sept.
  2016, pp. 11--15.

\bibitem{Cai2020SCLDGM}
S.~Cai, W.~Lin, X.~Yao, B.~Wei, and X.~Ma, ``Systematic convolutional low
  density generator matrix code,'' \emph{IEEE~Trans.~Inf.~Theory}, vol.~67,
  no.~6, pp. 3752--3764, Jun. 2021.

\bibitem{Kakhaki2012LDGM}
A.~M. Kakhaki, H.~K. Abadi, P.~Pad, H.~Saeedi, F.~Marvasti, and K.~Alishahi,
  ``Capacity achieving linear codes with random binary sparse generating
  matrices over the binary symmetric channel,'' in \emph{Proc.~
  Int.~Symp.~Inf.~Theor.~(ISIT)}, Jul. 2012, pp. 621--625.

\bibitem{Csizar2011}
I.~Csisz{\'{a}}r and J.~K{\"{o}}rner, \emph{Information Theory: Coding Theorems
  for Discrete Memoryless Systems~(Second edition)}.\hskip 1em plus 0.5em minus
  0.4em\relax New York, NY: Cambridge University Press, 2011.

\bibitem{slepian1973correlated}
D.~Slepian and J.~Wolf, ``Noiseless coding of correlated information sources,''
  \emph{IEEE Trans. Inf. Theory}, vol.~19, no.~4, pp. 471--480, Jul. 1973.

\bibitem{roth2006introduction}
R.~Roth, \emph{Introduction to Coding Theory}.\hskip 1em plus 0.5em minus
  0.4em\relax Cambridge University Press, 2006.

\bibitem{Litsyn2003}
S.~Litsyn and V.~Shevelev, ``Distance distributions in ensembles of irregular
  low-density parity-check codes,'' \emph{IEEE Trans. Inf. Theory}, vol.~49,
  no.~12, pp. 3140--3159, Dec. 2003.

\bibitem{1996Agrell}
E.~Agrell, ``Voronoi regions for binary linear block codes,''
  \emph{IEEE~Trans.~Inf.~Theory}, vol.~42, no.~1, pp. 310--316, Jan. 1996.

\bibitem{elias1957list}
P.~Elias, ``List decoding for noisy channels,'' \emph{IRE WESCON Conv. Rec.},
  vol.~2, pp. 94--104, 1957.

\bibitem{shannon1967lower}
C.~E. Shannon, R.~G. Gallager, and E.~R. Berlekamp, ``Lower bounds to error
  probability for coding on discrete memoryless channels.'' \emph{Inform.
  Contr.}, vol.~10, pp. 65--103~(Part I),~522--552 (Part II), 1967.

\bibitem{seshadri1994list}
N.~Seshadri and C.-E. Sundberg, ``List {Viterbi} decoding algorithms with
  applications,'' \emph{IEEE Trans. Commun.}, vol.~42, no. 234, pp. 313--323,
  Feb.-Apr. 1994.

\bibitem{1999Guruswami}
V.~Guruswami and M.~Sudan, ``Improved decoding of {Reed-Solomon} and
  algebraic-geometry codes,'' \emph{IEEE~Trans.~Inf.~Theory}, vol.~45, no.~6,
  pp. 1757--1767, Sept. 1999.

\bibitem{2012CRC}
K.~Niu and K.~Chen, ``{CRC}-aided decoding of polar codes,'' \emph{IEEE Commun.
  Lett.}, vol.~16, no.~10, pp. 1668--1671, Oct. 2012.

\bibitem{tal2015list}
I.~Tal and A.~Vardy, ``List decoding of polar codes,''
  \emph{IEEE~Trans.~Inf.~Theory}, vol.~61, no.~5, pp. 2213--2226, May. 2015.

\bibitem{Cover2006}
T.~M. Cover and J.~A. Thomas, \emph{Elements of Information Theory~(Second
  edition)}.\hskip 1em plus 0.5em minus 0.4em\relax John Wiley \& Sons, Inc.,
  Hoboken, New Jersey: Cambridge University Press, 2006.

\bibitem{Wang2022}
X.~Ma, Y.~Wang, and T.~Zhu, ``A new framework for proving coding theorems for
  linear codes,'' in \emph{Proc.~ Int.~Symp.~Inf.~Theor.~(ISIT)}, Jun. 2022,
  pp. 2768--2773.

\bibitem{Forney1991}
G.~Forney, ``Geometrically uniform codes,'' \emph{IEEE Trans. Inf. Theory},
  vol.~37, no.~5, pp. 1241--1260, Sept. 1991.

\bibitem{Gallager1963LDPC}
R.~G. Gallager, \emph{{Low-Density Parity-Check Codes}}.\hskip 1em plus 0.5em
  minus 0.4em\relax Cambridge, MA: MIT Press, 1963.

\bibitem{2013Ma}
X.~Ma, J.~Liu, and B.~Bai, ``New techniques for upper-bounding the {ML}
  decoding performance of binary linear codes,'' \emph{IEEE Trans. Commun.},
  vol.~61, no.~3, pp. 842--851, Mar. 2013.

\bibitem{Benedetto1996IRWEF}
S.~Benedetto and G.~Montorsi, ``Unveiling turbo codes: Some results on parallel
  concatenated coding schemes,'' \emph{IEEE Trans. Inf. Theory}, vol.~42,
  no.~2, pp. 409--428, Mar. 1996.

\bibitem{Ma2015BMST}
X.~Ma, C.~Liang, K.~Huang, and Q.~Zhuang, ``Block {Markov} superposition
  transmission: Construction of big convolutional codes from short codes,''
  \emph{IEEE~Trans.~Inf.~Theory}, vol.~61, no.~6, pp. 3150--3163, Jun. 2015.

\bibitem{2001EXIT}
S.~ten Brink, ``Convergence behavior of iteratively decoded parallel
  concatenated codes,'' \emph{IEEE Trans. Commun.}, vol.~49, no.~10, pp.
  1727--1737, Oct. 2001.

\bibitem{Newman2002PRL}
M.~E. Newman, ``Assortative mixing in networks,'' \emph{Phys.~Rev.~Lett.},
  vol.~89, no.~20, p. 208701, 2002.

\bibitem{Barabasi2016NS}
A.-L. Barab{\'a}si and M.~P{\'o}sfai, \emph{Network Science}.\hskip 1em plus
  0.5em minus 0.4em\relax Cambridge University Press, 2016.

\bibitem{Watts1998NAT}
D.~J. Watts and S.~H. Strogatz, ``Collective dynamics of `small-world'
  networks,'' \emph{Nature}, vol. 393, no. 6684, pp. 440--442, 1998.

\bibitem{Barabasi1999SCI}
A.-L. Barab{\'a}si and R.~Albert, ``Emergence of scaling in random networks,''
  \emph{Science}, vol. 286, no. 5439, pp. 509--512, 1999.

\bibitem{Pastor2001PRL}
R.~Pastor-Satorras and A.~Vespignani, ``Epidemic spreading in scale-free
  networks,'' \emph{Phys.~Rev.~Lett.}, vol.~86, no.~14, p. 3200, 2001.

\bibitem{Mezard2009information}
M.~Mezard and A.~Montanari, \emph{Information, physics, and computation}.\hskip
  1em plus 0.5em minus 0.4em\relax Oxford University Press, 2009.

\bibitem{1974BCJR}
L.~Bahl, J.~Cocke, F.~Jelinek, and J.~Raviv, ``Optimal decoding of linear codes
  for minimizing symbol error rate (corresp.),'' \emph{IEEE Trans. Inf.
  Theory}, vol.~20, no.~2, pp. 284--287, Mar. 1974.

\bibitem{2022FR}
S.~Cai, S.~Zhao, and X.~Ma, ``Free ride on {LDPC} coded transmission,''
  \emph{IEEE Trans. Inf. Theory}, vol.~68, no.~1, pp. 80--92, Jan. 2022.

\end{thebibliography}


%
%
%
%
%
%
%

\end{document}